\documentclass{emulateapj}
\usepackage{epsfig}

\newcommand\msun{\cal{M}_\odot}
\defcitealias{west10}{Paper I}

\shortauthors{Bochanski et al.}
\shorttitle{Statistical Parallax of SDSS Low--Mass Dwarfs}
\begin{document}

\title{The Sloan Digital Sky Survey DR7 Spectroscopic M Dwarf Catalog II: Statistical Parallax Analysis}

\slugcomment{Accepted by AJ}

\author{		John J. Bochanski\altaffilmark{1,2},	
                    Suzanne L. Hawley\altaffilmark{3},
		 Andrew A. West\altaffilmark{4,5}
	 }
\altaffiltext{1}{Astronomy and Astrophysics Department, Pennsylvania
  State University, 525 Davey Laboratory, University Park, PA 16802\\ 
email:jjb29@psu.edu}
\altaffiltext{2}{Kavli Institute for Astrophysics and Space Research, Massachusetts Institute of Technology, Building 37, 77 Massachusetts Avenue, Cambridge, MA 02139}
\altaffiltext{3}{Astronomy Department, University of Washington,
   Box 351580, Seattle, WA  98195}
\altaffiltext{4}{Department of Astronomy, Boston University, 725 Commonwealth Avenue, Boston, MA 02215}
\altaffiltext{5}{Visiting Investigator, Department of Terrestrial Magnetism, Carnegie
Institute of Washington, 5241 Broad Branch Road, NW, Washington, DC
20015}

\begin{abstract}
We present a statistical parallax analysis of low--mass dwarfs from
the Sloan Digital Sky Survey (SDSS).  We calculate absolute $r$-band
magnitudes ($M_r$) as a function of color and spectral type, and
investigate changes in $M_r$ with location in the Milky Way.  We find
that magnetically active M dwarfs are intrinsically brighter in $M_r$ than
their inactive counterparts at the same color or spectral type.  Metallicity, as traced by the proxy
$\zeta$,  also affects $M_r$, with metal poor stars having fainter
absolute magnitudes than higher metallicity M dwarfs at the same color
or spectral type.  Additionally, we measure the velocity ellipsoid and solar reflex motion for each subsample of M dwarfs.
We find good agreement between our measured solar peculiar motion and
previous results for similar populations, as well as some evidence
for differing motions of early and late M type populations in $U$ and $W$
velocities that cannot be attributed to asymmetric drift.  The reflex
solar motion and the velocity dispersions both show that younger populations,
as traced by magnetic activity and location near the Galactic plane, have
experienced less dynamical heating.  We introduce a new parameter, the
independent position altitude (IPA), to investigate populations as a function
of vertical height from the Galactic plane. M dwarfs at all types exhibit
an increase in velocity dispersion when analyzed in comparable IPA subgroups.

\end{abstract}

\section{Introduction}
Low--mass dwarfs (0.08 $\msun$ $< \cal{M} <$ 0.8 $\msun$) dominate the stellar content
of the Milky Way \citep{2010AJ....139.2679B}.  However, the study of these stars in large
numbers was only recently realized with the advent of
surveys such as the Sloan Digital Sky Survey \citep[SDSS;][]{2000AJ....120.1579Y}
and the Two Micron All Sky Survey
\citep[2MASS;][]{2006AJ....131.1163S}.  These surveys, with their
precise, multi--band photometry and (in the case of SDSS)
spectroscopic coverage have led to observational catalogs of
unprecedented size.  Currently, the largest spectroscopic database
of M dwarfs \citep[][hereafter Paper I]{west10} has produced spectral types,
radial velocities (RVs) and chromospheric activity estimates (as
traced by Balmer series emission) for over 70,000 M dwarfs.  SDSS
photometry of millions of stars has been used to measure the
field luminosity and mass functions of M dwarfs
\citep{covey08,2010AJ....139.2679B}, 
the structure of the Milky Way's thin and thick disks
\citep{2008ApJ...673..864J,2010AJ....139.2679B}, and the properties of
wide, common proper motion binaries \citep{2010AJ....139.2566D}.

The ubiquity and long main sequence lifetimes of low--mass stars \citep{1997ApJ...482..420L} make M dwarfs ideal tracers of
nearby Galactic structure and kinematics \citep[e.g.,][]{
  1993ApJ...409..635R,1995AJ....110.1838R,2007AJ....134.2418B, 2009AJ....137.4149F, 2010AJ....139.2679B}.  Despite the recent advances enabling the study of these
stars in a broad Galactic context, one fundamental parameter remains
elusive: distance.  Accurate trigonometric parallaxes, and the resulting absolute
magnitude and distance measurements, remain difficult to obtain due to the
intrinsic faintness ($L \lesssim 0.05 L_{\odot}$) of low--mass stars.  The
largest trigonometric parallax survey to date was performed by the
Hipparcos satellite \citep{1997yCat.1239....0E,2007A&A...474..653V}.  However, due to the relatively bright
magnitude limit of the Hipparcos sample ($V \lesssim 12$), M dwarfs observed
by Hipparcos are saturated in SDSS photometry, hindering the
construction of a color--absolute magnitude relation (CMR) in native
SDSS filters.  A few parallax estimates of SDSS M dwarfs have been
obtained \citep[e.g.,][]{2002AJ....124.1170D, 2004AJ....127.2948V}, but these measurements
are observationally taxing and take years to complete, and are thus
limited in number ($\sim$ 20).  Current (\citealp[][]{2010arXiv1008.0648R}; Faherty,
private communication) and future parallax studies, such as GAIA \citep{2001A&A...369..339P} will add valuable observations to this regime. 

Since trigonometric parallaxes for the vast majority of SDSS M dwarfs are
not available, alternate means to estimate their absolute
magnitudes must be employed.  \cite{2002AJ....123.3409H} and \cite{2005PASP..117..706W} used
SDSS--2MASS cross--matching to create $M_J$--spectral type relations,
and combined with average colors as a function of spectral type, to
construct SDSS CMRs.  Golimowski et al. (private communication) and
\cite{bochanskithesis} used $u^{\prime} g^{\prime} r^{\prime} i^{\prime} z^{\prime}$ photometry of $\sim 200$ nearby M dwarfs with accurate
trigonometric parallaxes, and transformed from the
$u^{\prime} g^{\prime} r^{\prime} i^{\prime} z^{\prime}$ system of the
SDSS photometric telescope to the
native SDSS $ugriz$ system using the relations of \cite{2007AJ....134.2430D}.
Other CMRs applicable for warmer stars in the SDSS filter--set have
been derived using
binaries \citep{Sesar08} and averages of existing CMRs
\citep{2008ApJ...673..864J}.  
Using a kinematic
 model for motions of stars in the Milky Way,
 \cite{2010ApJ...716....1B} derived a distance scale and CMR for main
 sequence stars observed by SDSS.  
Other studies have used 2MASS--SDSS
 color transformations \citep{2009MNRAS.396.1589B} and synthetic photometry
 \citep{2007AJ....134.2398C} to derive CMRs in the $ugriz$ filter system.

One method for constructing CMRs in the native SDSS system
that has not yet been utilized is the classical statistical parallax method.  Using proper
motions and radial velocities of a large number stars from a
homogeneous population, an estimate 
of the average absolute magnitude, the velocity ellipsoid, and the Sun's peculiar
motion can be obtained.   This method has a rich history in the
astronomical literature
\citep[e.g.,][]{1965BAN....18...71V,1971MNRAS.151..231C,1983veas.book.....M,1998ApJ...506..259P}, and has previously been
employed for RR Lyraes \citep[e.g.,][]{1986ApJ...302..626H,1986MNRAS.220..413S,1996AJ....112.2110L,1998A&A...330..515F,1998ApJ...506..259P} and Cepheids \citep[e.g.,][]{1991ApJ...378..708W}. 

In addition to absolute magnitude estimates, the statistical parallax
technique provides a measurement of the velocity ellipsoid of the
stellar population and the reflex motion of the Sun.  The velocity
ellipsoid is described by dispersions and directions along three
principal axes.  Low--mass stars make excellent tracers of the local
Galactic potential
\citep{1997PASP..109..559R,2007AJ....134.2418B,2009AJ....137.4149F} as constrained by the observed velocity dispersions.  The
solar peculiar motion determination complements studies using
Hipparcos proper motions of nearby, young stars \citep{Dehnen98}, SDSS
observations of M dwarfs \citep{2009AJ....137.4149F} and the
Palomar-Michigan State University survey of nearby M dwarfs \citep[PMSU;][]{1995AJ....110.1838R,1996AJ....112.2799H}.

In this paper, we use the large spectroscopic sample of M dwarfs described
in \citetalias{west10}
to determine absolute magnitudes in the native SDSS filters and examine kinematics using the statistical
parallax method.  The observations employed in our study are described
in \S \ref{sec:obs}.  We detail our statistical parallax analysis in
\S \ref{sec:method}.  Our results are presented in \S
\ref{sec:results}, with conclusions following in \S \ref{sec:conclusions}.

\section{Observations}\label{sec:obs}
Accurate and precise photometry, astrometry and velocities are required for
statistical parallax analysis.  The data were obtained from the 
latest SDSS data release \citep[DR7; ][]{2009ApJS..182..543A} which
contains photometry over nearly 10,000 sq.\ deg.\ down to faint magnitudes 
($r \sim$22) in five filters \citep[$ugriz$, ][]{1996AJ....111.1748F}.
When sky
conditions at Apache Point Observatory were not photometric, the SDSS
operated in a spectroscopic mode.  SDSS photometry was used to target objects for spectroscopic followup,
primarily galaxies
\citep{2002AJ....124.1810S} and quasars \citep{2002AJ....123.2945R}.
Approximately 460,000 stellar spectra were also obtained both as
targeted and serendipitous observations.  Twin fiber-fed spectrographs collected 640 observations
simultaneously, with individual 15--20--minute exposures being co-added
for a typical total exposure time of $\sim$ 45 minutes.  These
medium--resolution ($R \sim 2,000$) spectra cover the entire optical bandpass
(3800 - 9200 \AA).  

The absolute astrometric precision of SDSS is $< 0.1^{\prime\prime}$ in each
coordinate \citep{2003AJ....125.1559P}.  Proper motions for SDSS
objects were calculated by matching to the USNO-B survey
\citep{2004AJ....127.3034M,2008AJ....136..895M}.  The proper motions 
have a baseline of $\sim$ 50 years, and a typical precision
of $\sim$ 3 mas yr$^{-1}$ in right ascension and declination.

In \citetalias{west10}, we described the SDSS DR7 spectra that we compiled into
the largest spectroscopic catalog of M dwarfs ever constructed,
containing $\sim$ 70,000 stars.
Briefly, the catalog was obtained by color
selection of stars with $r-i > 0.42$ and $i-z > 0.24$, before 
correcting for Galactic reddening \citep{1998ApJ...500..525S}.  Each star was processed with the Hammer IDL
package \citep{2007AJ....134.2398C}.  The Hammer provides spectral
type estimates and measures a number of spectral features, including
H$\alpha$ equivalent width and various TiO  and CaH bandhead strengths.  Each spectrum
was then visually inspected and non-M dwarf contaminants were culled
from the catalog.  Spectral types were determined by comparing to the
SDSS templates \citep{2007AJ....133..531B}.  The RVs were measured by cross--correlating
each spectrum with the appropriate template from 
\cite{2007AJ....133..531B}, with a precision of $\sim$ 7 km s$^{-1}$.
See \citetalias{west10} for more detailed information about the selection and
characteristics of the DR7 catalog.

For the sample used in the statistical parallax analysis, 
we required that stars have well--measured proper motions and radial
velocities, and precise photometry 
(see \cite{2010AJ....139.2679B} for information on photometric flag
cuts, and \cite{2010AJ....139.2566D} for our kinematic quality flags).  
Our quality cuts reduced the original sample to 40,963 stars.  

\section{Method}\label{sec:method}
We employed the maximum likelihood formulation of classical 
statistical parallax analysis as presented by \cite{1983veas.book.....M}
and used previously to analyze RR Lyrae stars 
\citep{1986ApJ...302..626H,1986MNRAS.220..413S,1996AJ....112.2110L,1998A&A...330..515F} and Cepheids \citep{1991ApJ...378..708W}.  A variation
of the model was used to investigate the kinematics of the nearby
M dwarfs from the PMSU 
sample \citep{1996AJ....112.2799H}.  Briefly, this statistical
parallax method models the velocity
distribution of a homogeneous stellar 
population with nine kinematic parameters,
including the three velocities of the reflex solar motion,
and the three directions and three dispersions of the
velocity ellipsoid, which describes the random and peculiar
velocities of the population.  The stars are assumed
to have an absolute magnitude with some intrinsic
dispersion, and these two
additional parameters determine the distance used
to transform the observed proper motions into transverse 
velocities.  The eleven parameters used in the model are described in Table
\ref{table:parameters}. 

The observational data needed for the analysis are
the position, apparent magnitude (corrected for Galactic extinction),
proper motions and radial velocity for each star in the
sample.  The uncertainties in the data are used to assign
appropriate weights in the solution.  
We solved for the eleven parameters in the model
simultaneously by maximizing the likelihood using
geometric simplex optimization \citep{optimizationsimplex_nelder_1965,1978Daniels}.  Uncertainties in the parameters were
determined by numerical computation of the derivative
of each parameter individually, while keeping all other
parameters fixed.  The maximum likelihood equations and 
simplex method are described in detail in \cite{1986ApJ...302..626H}.

With any multi--parameter fit, sensible
constraints are necessary to ensure the model does not prefer an
unrealistic portion of $\chi^2$ space.  For this analysis, we fixed one parameter, $\sigma_k$.  This parameter is
related to the spread in absolute magnitude, $\sigma_M$ by the
following equation from \cite{1986ApJ...302..626H}:
\begin{equation}
{\sigma_M}^2 = \log_{10}[1 + {\sigma_k}^2/(1 + k)^2]/(0.04~\ln10)
\label{eqn:sigk}
\end{equation}
where $\sigma_M$ is the spread in absolute magnitude for a given
luminosity bin, and $\sigma_k$ is the spread in $k$ which is a distance
scale parameter.  The absolute magnitude is related to $k$ through the following equation:
\begin{equation}
{M_r} = 5\log_{10}(1 + k) + M_A - 0.1~\ln10 (\sigma_M)^2
\end{equation}
where $M_A$ is an initial estimate of the absolute magnitude.  We fixed $\sigma_k$ at four
values: 0.05, 0.1, 0.2 and 0.3.   Usually $k \sim 0.01-0.02$, so
$\sigma_k = 0.2$ corresponds to $\sigma_M \sim 0.4$ which is the
typical dispersion in absolute magnitude for low--mass stars \citep{2010AJ....139.2679B}.

\begin{center}
\begin{deluxetable}{lrr}
\tablewidth{0pt}
\tabletypesize{\small}
 \tablecaption{Statistical Parallax Fit Parameters}
 \tablehead{
 \colhead{Parameter} &
 \colhead{Units} &
 \colhead{Description} 
}
 \startdata
$\sigma_{U}$ & km s$^{-1}$ & Velocity dispersion in radial direction\\
$\sigma_{V}$ & km s$^{-1}$ & Velocity dispersion in orbital direction\\
$\sigma_{W}$ & km s$^{-1}$ & Velocity dispersion in vertical direction\\
\bf{$r,\phi,z$} & radians & Orientation of velocity ellipsoid\\
$U$ & km s$^{-1}$ & Solar peculiar motion ($r$)\\
$V$  & km s$^{-1}$& Solar peculiar motion ($\phi$)\\
$W$ &km s$^{-1}$ & Solar peculiar motion ($z$)\\
$k$ & \nodata & Distance scale\\
$\sigma_k$ & \nodata & Dispersion in distance scale\\

\enddata
 \label{table:parameters}
\end{deluxetable}
\end{center}

\subsection{Constructing Subsamples for Analysis}
Our spectroscopic sample is much larger
than was available for previous statistical parallax studies, 
which contained
observations of a few hundred or less objects.  Thus, we were 
able to divide the
sample into smaller subsamples, selected on color, spectral type,
magnetic activity (as traced by H$\alpha$), metallicity (using the
$\zeta$ parameter of \citealp{2007ApJ...669.1235L}), position on the sky
(northern and southern Galactic hemispheres), distance (estimated
from the $M_r,~r-z$ CMR relation from \citealp{2010AJ....139.2679B}),
and Galactic height (discussed further below).
We required
at least 100 stars in each subsample to ensure a meaningful solution.  
Table  \ref{table:subsamples} 
lists the properties of the various subsamples.

The
color, spectral type, metallicity and magnetic activity subsamples were
chosen to explore intrinsic variations within the M
dwarf population, as traced by their absolute magnitudes and kinematics.  The
position and distance subsets were examined to ensure that we
were not biased by extinction or proper motion limits.  While
the majority of our sample is contained at high Galactic latitudes ($b
\sim 50^{\circ}$) in low extinction regions (the
median extinction in $r$ is 0.07 mags), some sightlines to the south (through the Plane)
may be affected by extinction and reddening.  Current efforts to model
the 3-D dust distribution within the SDSS footprint (Jones et al.,
private communication) will improve the extinction and reddening
estimates for these stars.

\begin{figure}[htbp]
\centering
\includegraphics[scale=0.4]{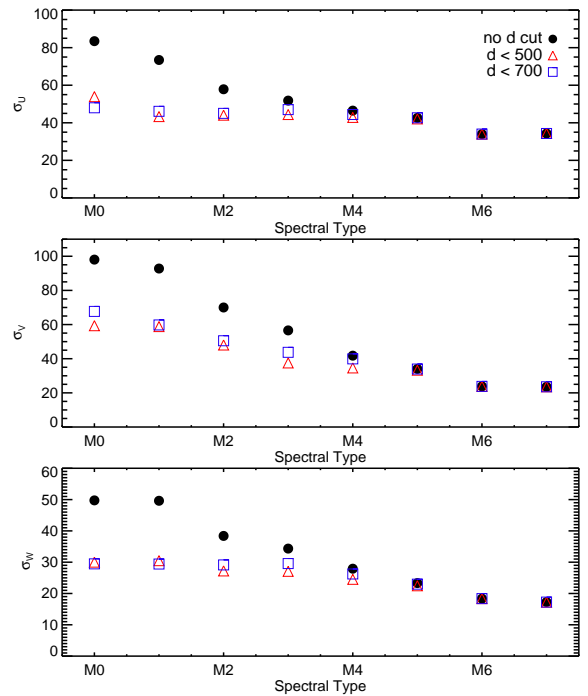} 

 \caption{$UVW$ dispersions vs. spectral type for $d < 500$ pc (open triangles), $d < 700$ pc (open squares) and no distance
   limit (filled circles).  Including more
   distant stars beyond our proper motion precision limit artificially
   inflates the measured dispersions for early--type stars.}
 
  \label{fig:dists_disps}
\end{figure}

Our proper motion catalog, described in \cite{2004AJ....127.3034M,2008AJ....136..895M} has a precision limit of 3 mas yr$^{-1}$.
For a typical thin disk star with a transverse velocity of $\sim$ 10
km s$^{-1}$, this limit corresponds to a limiting distance of $\sim$
700 pc.  Beyond this distance, only stars with larger transverse
motions would be measured with precise proper motions, artificially
biasing the measured velocity dispersion.  Thus, two distance limits
were explored:  stars with $d < 500$ pc and $d < 700$ pc.  The 500 pc
cut corresponds to a more conservative estimate of the proper motion
precision.  The effects of these distance limits are shown in Figure
\ref{fig:dists_disps}.  Early--type M dwarfs, which are found at
larger distances, are biased towards larger velocity dispersions if
a distance limit is not enforced.  The 500 pc and 700 pc limits give
similar results, so we adopt a $d < 700$ pc
limit for the remainder of this study.  Further kinematic analysis is
discussed in \S \ref{sec:kinematics}.  The $d < 700$ pc cut limits the
total number of stars in our sample to 22,542.  In Table
\ref{table:heights} we report the median, minimum and maximum heights
above the Plane for the stars in our $d < 700$ pc sample.  
The heights were measured using distances from the  $M_r, r-z$ 
color--magnitude relation of \cite{2010AJ....139.2679B}.

Finally, to probe changes in the kinematics and absolute magnitude as
a function of height in the Milky Way disk, we divided each spectral type
and color bin according to the independent position altitude (IPA).  At a
given spectral type, bins in apparent magnitude will correspond to
spherical shells in space.  The IPA cuts are
slices in Galactic height, defined as:
\begin{equation}
{\rm IPA} \equiv 10^{m_r/5}  \sin b
\label{eqn:ipa}
\end{equation}
where $m_r$ is the observed $r$--band magnitude and $b$ is the Galactic
latitude of the star.  We stress that the IPA can only be used for
relative comparisons within the same intrinsic luminosity bins.  That
is, M0 and M6 stars with the same IPA value
will be located at very different Galactic heights.  The IPA is useful as
it avoids the assumption of an absolute magnitude, allowing an independent
determination for each IPA subset.  It is based entirely
on observable properties, and is somewhat analogous to the often--used reduced proper motion \citep{1925ApJ....62....8L}.

\begin{center}
\begin{deluxetable*}{llll}
\tablewidth{0pt}
\tabletypesize{\small} 
\tablecaption{SDSS Subsamples}
 \tablehead{
 \colhead{Initial Cut} &
 \colhead{Bin Size} &
 \colhead{Further Criteria} &
 \colhead{Comments} 
}
 \startdata
Spectral Type &  1 type bins & \nodata  & \nodata \\   %8 bins
Color             &   0.3 mag bins & \nodata & \nodata \\
                  &    &   Activity & H$\alpha$ active stars defined in \citetalias{west10}  \\ %8 + 8 bins
                  &    & Hemisphere & N and S Galactic Hemispheres   \\  % 8 + 7 bins
                  &    & $\zeta$   & 0.25 bins       \\
                  &    & IPA        &  10 bins per spectral type \\  % 72 bins
                 &     & Distance & $< 500$ pc \& $< 700$ pc\\ % 8 + 8 bins
\enddata
 \label{table:subsamples}
\tablecomments{Bins with fewer than 100 stars were not used in this analysis.}
\end{deluxetable*}
\end{center}

\begin{center}
\begin{deluxetable*}{lllll}
\tablewidth{0pt}
\tabletypesize{\small} 
\tablecaption{Galactic Heights of SDSS M Dwarfs}
 \tablehead{
 \colhead{} &
 \colhead{Median Height} &
 \colhead{Min.  Height} &
 \colhead{Max Height} &
 \colhead{$N_{\rm Stars}$}
}
 \startdata
Spectral Type &  &  & & \\   %8 bins
\hline
M0 &    395   &     20   &    690   &   1323  \\
M1 &    392   &     22   &    689   &   1877  \\
M2 &    337   &     11   &    693   &   4700  \\
M3 &    274   &     14   &    712   &   6156  \\
M4 &    214   &      9   &    694   &   4784  \\
M5 &    137   &      9   &    694   &   1616  \\
M6 &    108   &      4   &    423   &   1333  \\
M7 &     82   &      0   &    230   &    753  \\
\hline
$r-z$ &  &  & & \\   %8 bins
\hline
0.75 &    402   &     20   &    690   &    348  \\
1.05 &    395   &     23   &    688   &   1647  \\
1.35 &    375   &     16   &    693   &   2914  \\
1.65 &    309   &     11   &    712   &   5937  \\
1.95 &    254   &     13   &    694   &   5639  \\
2.25 &    181   &      9   &    634   &   2922  \\
2.55 &    123   &     11   &    387   &    818  \\
2.85 &    118   &      7   &    304   &    958  \\
3.15 &     98   &      0   &    221   &    960  \\
3.45 &     76   &      2   &    340   &    296  \\
3.75 &     58   &      5   &    126   &    102  \\

\enddata
\label{table:heights}
\tablecomments{All heights reported in pc and measured using distances from the
  $M_r, r-z$ CMR of \cite{2010AJ....139.2679B}.}
\end{deluxetable*}
\end{center}

\subsection{Computational Method}
Following the method of \cite{1986ApJ...302..626H} each subsample of
M dwarfs was analyzed using the following prescription.  Ten
runs were calculated for each dataset.  The initial absolute
magnitude estimates were derived from the $M_r, r-z$ CMR of
\cite{2010AJ....139.2679B}.  The absolute
magnitude estimate was updated after each run, with the output of the
previous run being used as input for the next.  
For a given run, the simplex optimization iterated 5,000 times.
Thus, for each subsample, 50,000 iterations were computed.  This ensured
that the simplex operator had sufficient freedom to explore parameter
space and converged to the best (maximum likelihood) answer.  Typically,  
convergence was obtained after 25,000 iterations.  

\section{Results}\label{sec:results}
The results of our analysis are detailed below.  The outputs from the
code returned three major results:  the absolute magnitude of a given
stellar sample, the Sun's velocity with respect to that sample, and
the velocity ellipsoid that describes the peculiar motions of the sample.  Each of these results
are examined, and compared to previous investigations.
In the discussion, we use the $\sigma_k = 0.2$ results as our
fiducial value, since it
corresponds to $\sigma_{M_r} = 0.4$ which is observed for nearby stars
with trigonometric parallaxes in the $M_r, r-z$ color--magnitude diagram.

\subsection{Absolute Magnitudes}
Traditionally, colors or spectral types have been employed to
estimate absolute magnitude when trigonometric parallaxes were not
available.  These are often referred to as photometric and
spectroscopic parallaxes, respectively.  In Figure
\ref{fig:mr_sp_rz}, we show the spectroscopic (left panel) and
photometric (right panel) parallax relations for
M dwarfs, compared to previous studies in the $ugriz$ system
\citep{2002AJ....123.3409H, 2005PASP..117..706W, 2007ApJ...662..413K}.   The open circles are nearby low--mass stars
with accurate trigonometric parallaxes, tabulated in
\cite{bochanskithesis}.  Using the Hammer \citep{2007AJ....134.2398C}, we manually
assigned a spectral type to each nearby star, using archival spectra from
multiple sources:  the PMSU study \citep{1995AJ....110.1838R,1996AJ....112.2799H}, SDSS,
the local star sample of \cite{2002AJ....123.2828C} and the
DwarfArchives website\footnote{Available at \url{http://www.dwarfarchives.org}}.  While many of the spectral
types agreed with previous results, we did find some disagreements
at the level of $\pm 2$ spectral classes.  Therefore, we include
updated spectral types for these nearby stars in Table
\ref{table:spec_types}.  The left panel of Figure \ref{fig:mr_sp_rz}
highlights the large dispersion in $M_r$ as a function of spectral
type.  The typical spread is $\sim 0.5$ mag, increasing 
to $\sim 0.8$ magnitudes near type M4, nearly double the spread for a
given $r-z$ color bin.  Thus, we strongly suggest that spectral type
should \textit{not} be used as a tracer of absolute magnitude for M
dwarfs.   The use of half--interger spectral types may decrease the
observed scatter in Figure \ref{fig:mr_sp_rz}.   However,
half--integer spectral types would require
higher resolution and higher S/N spectra, which is not practical for this
type of large survey sample.  Furthermore, half--interger spectral
type standards are not defined for the entire M dwarf sequence
\citep{1991ApJS...77..417K}.  Finally, half--integer types will likely
span a larger color range than the color bins we have used (see
\citetalias{west10} for a discussion of spectral types and mean
colors).  

The left panel of Figure \ref{fig:mr_sp_rz} also displays disagreement between our
statistical parallax results (solid black, red and blue lines and
filled circles) and the nearby star sample at later
types.  This is due to a systematic color difference within a given
spectral type bin between the two samples.  That is, the nearby star
sample is systematically redder than the SDSS stars at a given
spectral type, possibly due to the small number of nearby stars in 
these bins.  
 Thus, the SDSS stars have an absolute magnitude (at a
given spectral type) that is appropriate for the bluer, more luminous
stars of that type.  This effect disappears when $r-z$ color is substituted 
for spectral type (Figure \ref{fig:mr_sp_rz}, right panel).  

The right panel of Figure \ref{fig:mr_sp_rz} shows our statistical parallax $M_r, r-z$ relation, along
with the results of previous studies \citep{2002AJ....123.3409H, 2005PASP..117..706W, 2010AJ....139.2679B}. Overplotted with open
circles are the absolute magnitudes and colors from the nearby star
sample listed in Table \ref{table:spec_types}.  Note that the
dispersion in absolute magnitude as a function of color is
significantly smaller than that shown in left panel of Figure \ref{fig:mr_sp_rz},
indicating that the $r-z$ color is a much better tracer of absolute
magnitude.  Furthermore, the discrepancy in mean color for a given
spectral type between the nearby star sample and SDSS stars is not
evident.  We reiterate that colors, rather than spectral types, 
are preferred for absolute magnitude and hence distance estimation
for M dwarfs.  

\subsubsection{Absolute Magnitude Variations and Magnetic Activity}
Due to the large spread in $M_r$ as a function of spectral type, we
focus our absolute magnitude analysis on our color subsamples (i.e.,
the right panel of Figure \ref{fig:mr_sp_rz}).
Our results are shown in the solid black, red and blue lines and
filled circles in Figure \ref{fig:mr_sp_rz}.  There are a few notable trends.  
First, the
bifurcated main sequence at blue colors (earlier spectral types) indicates that
active, early--type M dwarfs are intrinsically more luminous than their inactive
counterparts.  Recent observations of eclipsing binaries
\citep{2007ApJ...660..732L,2010ApJ...718..502M} and active, single stars \citep{2006ApJ...644..475B,2008A&A...478..507M}
suggest that active stars possess larger radii than inactive
dwarfs.  However, possible changes in effective temperature and
luminosity have not been well constrained.  Models
predict both unchanged \citep{2007A&A...472L..17C} or decreased
\citep{2001ApJ...559..353M} luminosity, while some observations
suggest that active low--mass dwarfs are over-luminous compared to their
model predictions \citep{2009ApJ...697..713M, 2010MNRAS.tmp.1180S}.  A
similar effect was observed in the PMSU sample, where active M dwarfs
were brighter in $V$ than inactive stars (Figure 4 of \citealp{1996AJ....112.2799H}).
The behavior evidenced in Figure \ref{fig:mr_sp_rz} suggests that
active, early--type M dwarfs are $\sim 1$ mag brighter in $M_r$, 
which could be explained if they have larger radii than inactive
stars at the same spectral type or color.  Direct
interferometric measurements of stellar radii are necessary to
independently quantify the amount the radius changes for 
an individual star, since our
statistical analysis is performed on subsamples of stars that
span a (small but nonzero) range of temperature.
We note that many close binaries are unresolved in
SDSS photometry.  Binarity may increase activity in early 
M dwarfs in close orbits, and could explain some of the absolute magnitude 
differences for those bins.   

At blue colors ($r-z < 2$), the inactive and total sample absolute
magnitudes fall below the mean locus of nearby trigonometric parallax
stars.  Since the
nearby stars are composed of a mix of active and inactive stars,
activity is likely not the only important effect.
We therefore next investigate the effects of metallicity on $M_r$ and attempt
to isolate them from those traced by chromospheric activity.

\subsubsection{Absolute Magnitude Variations and Metallicity}
While magnetic activity may affect the radius of a star, metallicity
can alter the star's effective temperature and luminosity, manifesting
as a change in absolute magnitude.
At a given color (or spectral type), stars with lower
metallicities will exhibit fainter absolute magnitudes
\citep{1959MNRAS.119..278S,2008ApJS..179..326A}\footnote{Alternatively,
  at a given mass, stars with lower metallicities are
  bluer (hotter) and more luminous than their high metallically counterparts.}.  
Without
precise metallicities and trigonometric parallaxes for all the M
dwarfs in our sample, we cannot directly examine this effect.  There
are some promising metallicity estimators for low--mass stars being
developed in the 
near-IR \citep{2009ApJ...699..933J, 2010ApJ...720L.113R}, but 
these are not applicable to SDSS spectroscopy.

We therefore investigated the effects of metallicity 
using the $\zeta$ parameter, as defined by
\cite{2007ApJ...669.1235L}.  $\zeta$ is a relative metallicity proxy
which uses the relative strengths of CaH and TiO bands to discriminate 
between stars of different composition.  It can
be employed as a rough tracer of $[Fe/H]$, with $\zeta = 1$ corresponding
to solar metallicity and $\zeta = 0.4$ corresponding to $[Fe/H] \sim
-1$ \citep{2009PASP..121..117W}.  We note that the $[Fe/H], \zeta$
relation is only calibrated over spectral types M0-M3 and suffers from
large spreads near solar metallicity.
In Figure \ref{fig:mr_sp_rz_zeta}, we plot the absolute
magnitudes for two bins in
$\zeta$ as a function of spectral type (left panel) and $r-z$ color
(right panel) for active (solid lines) and inactive (dashed lines) stars.  
Table \ref{table:abs_mags} gives the $M_r$ results for the various
$r-z$, activity and $\zeta$ bins.  The spread in each bin is $\sigma \approx 0.4$ mags,
set by the choice of $\sigma_{k} =$ 0.2.  As with any magnitude
limited survey (such as SDSS), \cite{m36} bias can arise.  For surveys with
complicated selection effects, a magnitude--independent
correction is preferred \citep{1998ApJ...506..259P}.  For an
exponential Galactic disk and $\sigma_{M_r} = 0.4$, the Malmquist correction is
0.15 mags \citep{2005nlds.book.....R}.  This correction has been applied to the values reported in
Table \ref{table:abs_mags}, and is much smaller than the reported uncertainty.

\begin{figure*}[htbp]
  \centering
 \begin{center}$
\begin{array}{cc}  
  \includegraphics[scale=0.4]{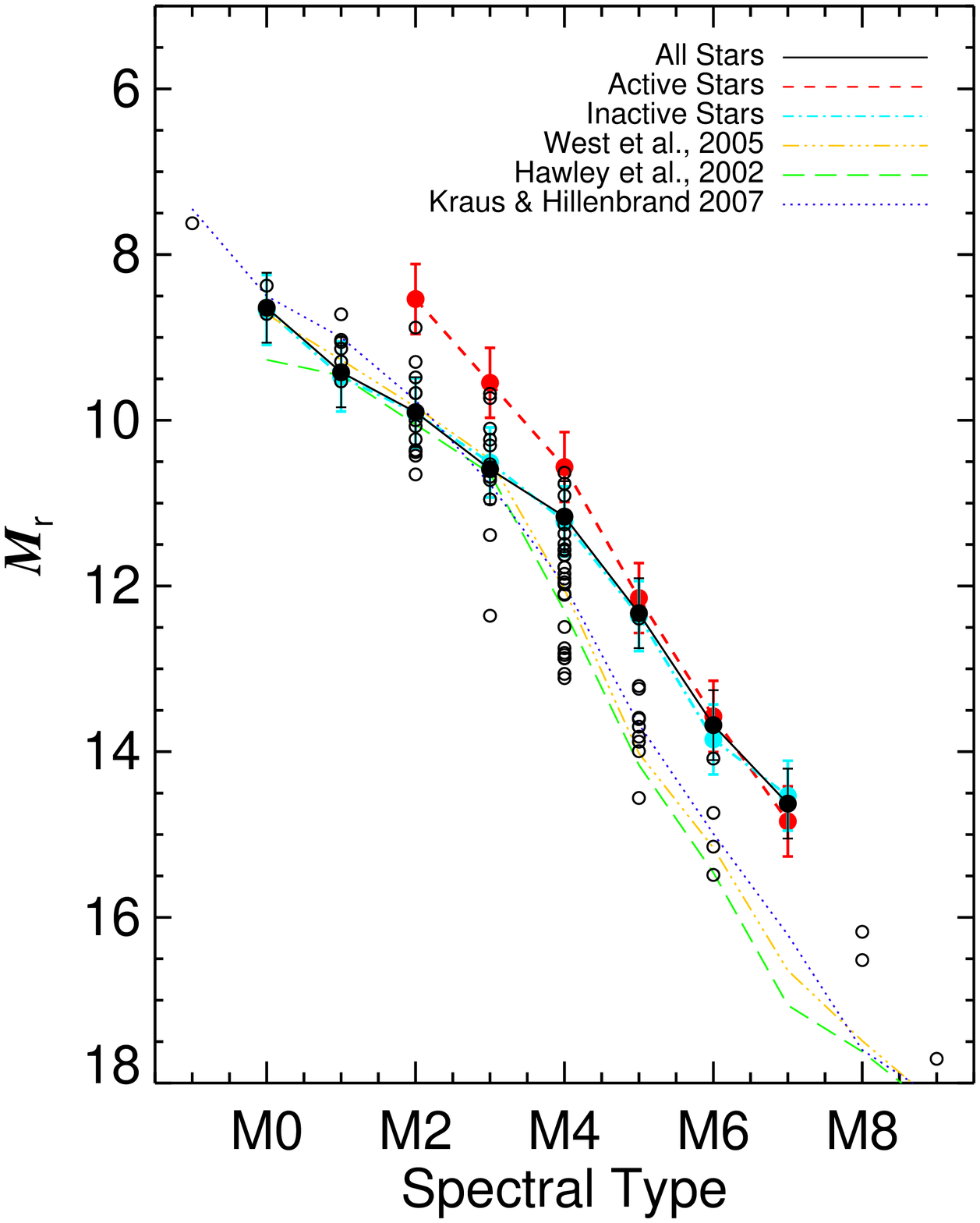}&
  \includegraphics[scale=0.4]{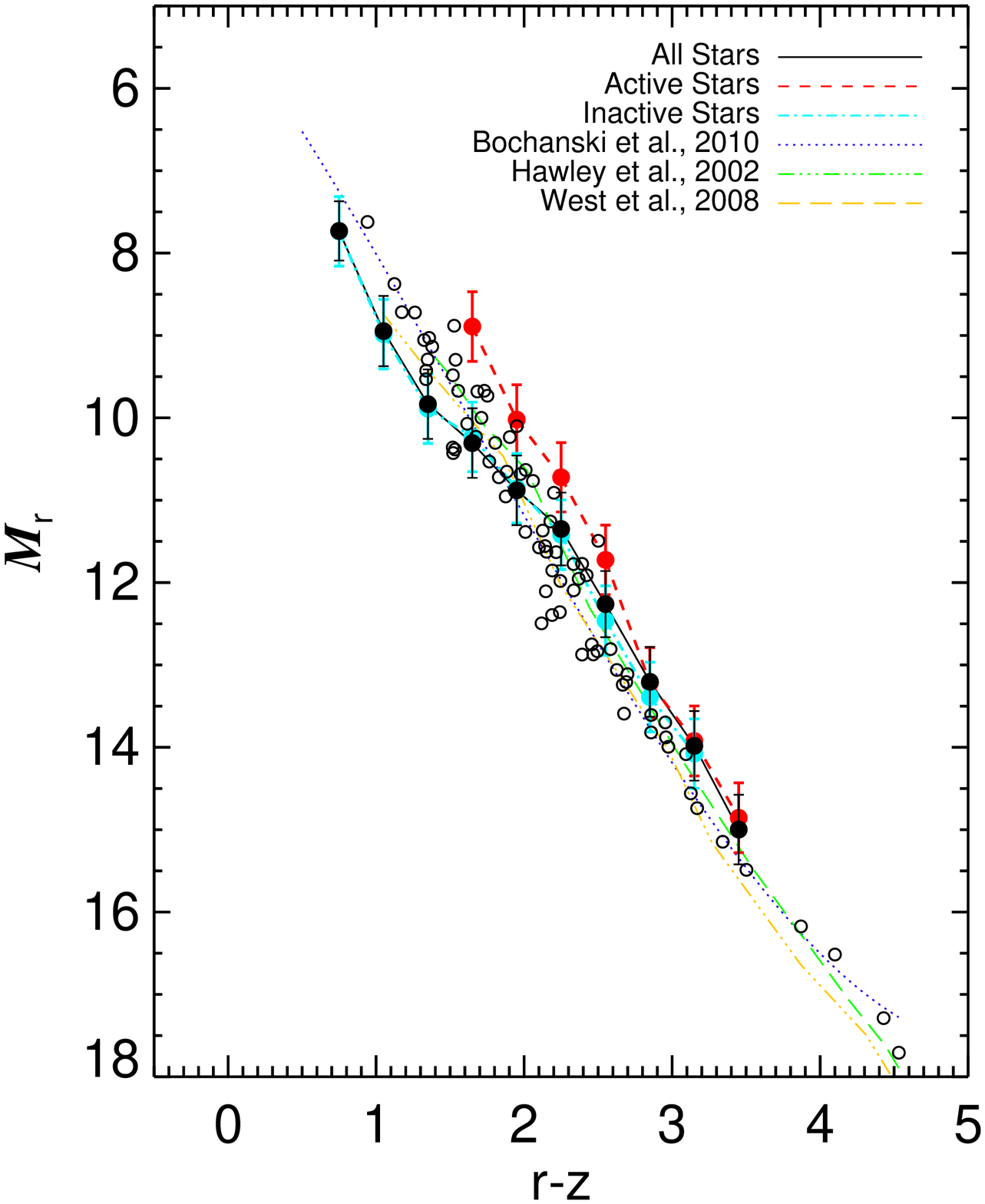} 
\end{array}$
\end{center}  
\caption{Left Panel: $M_r$ vs. spectral type.   The open circles are nearby
    stars with accurate trigonometric parallaxes from
    \cite{bochanskithesis}.  Previous spectroscopic parallax
    relations are overplotted and described in the legend.  The
    results from this study (black, red and blue lines and black filled
    circles) are plotted for comparison.  Note the large dispersion in
  $M_r$ for a given spectral type, especially at M4 and M5.  We
  strongly suggest that spectral type should not be used to estimate
  absolute magnitude in the SDSS $r$--band.
  Right Panel: $M_r$ vs. $r-z$, with same symbols and lines as in
the spectral type panel.  Previous studies are shown and described 
in the legend.  There is a much smaller dispersion in $M_r$ 
at a given $r-z$ color compared to the spectral type relations.}
  \label{fig:mr_sp_rz}
\end{figure*}

\begin{figure*}[htbp]
 \begin{center}$
\begin{array}{cc} 
 \includegraphics[scale=0.4]{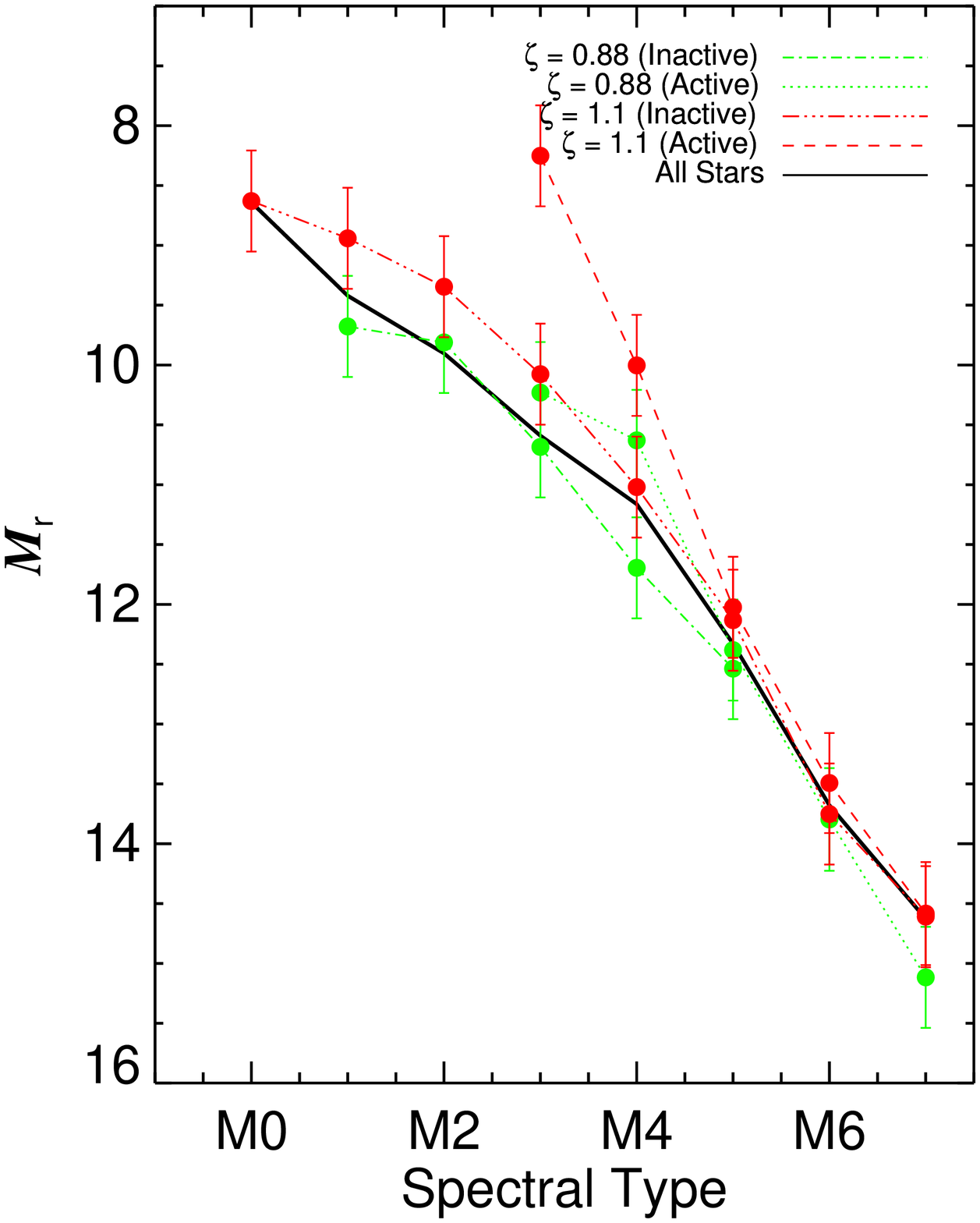} &
\includegraphics[scale=0.4]{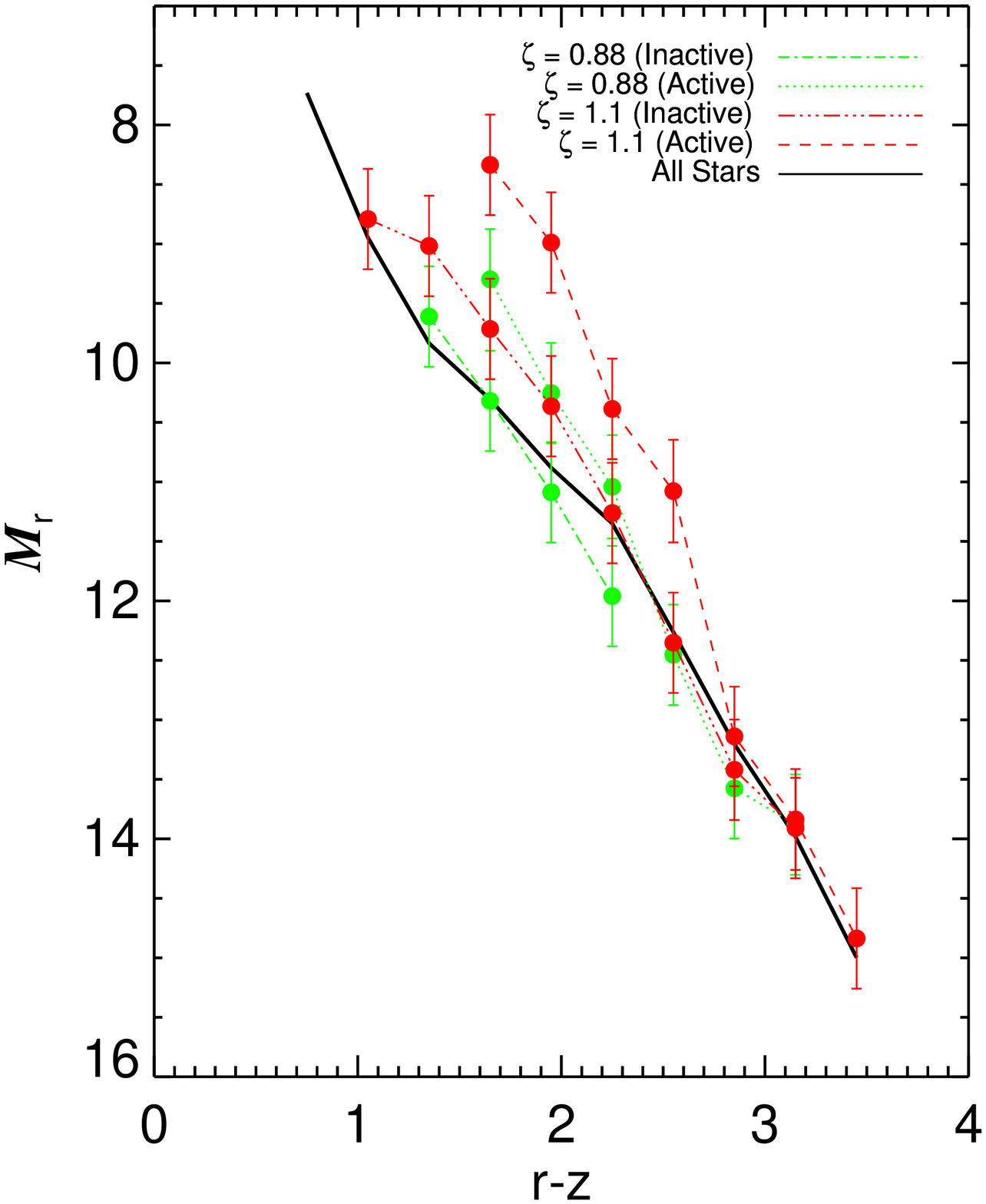}
\end{array}$
\end{center}   
\caption{Left Panel:  $M_r$ vs. spectral type for two values of
  $\zeta$, a metallicity proxy. Active stars (dashed and dotted lines) and
  inactive stars (dot--dash lines) at the same metallicity are plotted.   
   The total sample is overplotted in the solid black line. 
At a given
metallicity, active stars are brighter than their inactive
counterparts, while low--metallicity stars ($\zeta=0.88$) are dimmer than
higher metallicity M dwarfs ($\zeta=1.1$) at the same spectral type and activity
state.  Right Panel:  The $M_r$ vs. $r-z$ results show the same features
as in the spectral type panel.}
  \label{fig:mr_sp_rz_zeta}
\end{figure*}

Figure \ref{fig:mr_sp_rz_zeta} demonstrates that for stars with
similar chromospheric properties, the lower metallicity ones have 
fainter absolute
magnitudes, at the same color or spectral type. 
This is consistent with cluster studies within the SDSS
footprint \citep{2008ApJS..179..326A} and trigonometric parallax
studies of subdwarfs \citep{1997AJ....114..161R}.  Active stars at the
same metallicity are brighter than their inactive counterparts, but
Figure \ref{fig:mr_sp_rz_zeta} shows that both metallicity and
activity are important for determining the luminosity of an individual
star.
At a fixed $\zeta$,
the difference in $M_r$ is consistent with Figure \ref{fig:mr_sp_rz},
with early--type active stars being $\sim 1$ mag brighter, and the
difference diminishing at later spectral types.  A similar offset was
also measured for the same stars in $M_g$.
The total SDSS sample is
overplotted in each figure with a solid black line.  The
early--type stars, which are seen at larger distances due to SDSS
magnitude limits (see Table \ref{table:heights}), fall 
near the lower metallicity, inactive loci, while the
later type stars, seen closer to the Sun, are consistent with solar
metallicities and active stars.  This suggests that the low--mass dwarfs 
are tracing
a metallicity gradient similar to the one observed in SDSS
observations of higher--mass stars \citep{Ivezic08}, and also
explains why the SDSS sample falls below the locus of nearby stars
at early types (right panel of Figure \ref{fig:mr_sp_rz}).

We also note that the activity--metallicity loci appear to converge
near M5 ($r-z \sim 2.8$).  This behavior is not well explained, but
may be linked to the transition between a partially and fully 
convective stellar
interior that occurs near that spectral type/color.
Perhaps this transition, which 
alters the efficiency of energy transport in the star, also regulates
the luminosity at the surface.

Finally, we examined the IPA and hemisphere subsamples.  
No significant differences were observed in the
spectroscopic or photometric parallax relations with respect to
position in the Galaxy.  A trend with IPA was expected, since this
proxy for height probably traces metallicity.  The lack of a gradient is probably due to stars at various
heights being scattered into the same IPA bin due to photometric
errors.  The agreement between the hemisphere subsamples indicates
that the extinction correction is adequate for the sample.

%\pagebreak
\LongTables
\begin{center}
\begin{deluxetable}{lllll}
\tablewidth{0pt}
 \tablecaption{Nearby Star Sample Spectral Types}
 \tabletypesize{\scriptsize}

 \tablehead{
 \colhead{Name} &
 \colhead{$\alpha$\tablenotemark{a}} &
 \colhead{$\delta$\tablenotemark{a}} &
 \colhead{Sp. Type} &
 \colhead{Source\tablenotemark{b}}
}
 \startdata
GJ1002     &   1.67972278 &    -7.538550 & M6 & 1   \\
GJ1025     &  15.23465440 &    -4.449231 & M5 & 1   \\
Gl54.1     &  18.12784543 &   -16.998746 & M5 & 1   \\
LHS1302    &  27.76670103 &    -6.117972 & M5 & 1   \\
LHS1326    &  30.56747119 &    10.337209 & M6 & 1   \\
LHS1358    &  33.22773200 &     0.004801 & M4 & 1   \\
LHS1375    &  34.12457669 &    13.586802 & M6 & 1   \\
GL109      &  41.06474242 &    25.523168 & M2 & 1   \\
T832-10443 &  43.10949326 &     0.939522 & M9 & 3   \\
LHS168     &  48.34574221 &     4.774824 & M5 & 1   \\
GJ1065     &  57.68434297 &    -6.095079 & M3 & 1   \\
Gl169.1A   &  67.79870688 &    58.976686 & M4 & 1   \\
LHS1723    &  75.48913133 &    -6.946128 & M4 & 1   \\
G99-49     &  90.01463326 &     2.706515 & M4 & 1   \\
LHS1809    &  90.62144398 &    49.865368 & M5 & 1   \\
Gl232      &  96.17218963 &    23.432802 & M4 & 1   \\
Gl251      & 103.70407576 &    33.268001 & M3 & 1   \\
GJ1093     & 104.87030750 &    19.348539 & M5 & 1   \\
BL Lyn      & 112.98873721 &    36.229615 & M4 & 1   \\
2M0746+20  & 116.67677307 &    20.008842 & L0 & 4   \\
GJ1105     & 119.55303380 &    41.303596 & M4 & 1   \\
GJ2066     & 124.03289059 &     1.302435 & M2 & 1   \\
GJ1111     & 127.45563147 &    26.775994 & M8 & 1   \\
T213-2005  & 155.36376953 &    50.917938 & L0 & 4   \\
LHS283     & 158.86050669 &    69.449381 & M4 & 1   \\
Gl445      & 176.92332403 &    78.691261 & M4 & 1   \\
Gl447      & 176.93515470 &     0.804025 & M4 & 1   \\
GJ1151     & 177.73956422 &    48.377048 & M4 & 1   \\
GJ3693     & 178.46963501 &     6.998430 & M8 & 2   \\
Gl452.4    & 178.73946569 &    28.737442 & K7 & 1   \\
Gl455      & 180.57503136 &    28.587004 & M3 & 1   \\
GJ1156     & 184.74706541 &    11.126184 & M6 & 1   \\
Gl463      & 185.75015884 &    64.030910 & M3 & 1   \\
GJ1159A    & 187.30915386 &    53.545801 & M4 & 1   \\
LHS2633    & 191.75375224 &    46.625855 & M2 & 1   \\
Gl493.1    & 195.13932515 &     5.685628 & M5 & 1   \\
Gl514      & 202.49948063 &    10.376790 & M1 & 1   \\
Gl521      & 204.85046387 &    46.186687 & M1 & 1   \\
LHS2784    & 205.68017011 &    33.289742 & M4 & 1   \\
LHS2884    & 213.82055327 &    45.014702 & M3 & 1   \\
GJ3849     & 217.17973328 &    33.176899 & L0 & 4   \\
Gl552      & 217.37332947 &    15.533145 & M2 & 1   \\
GJ3855     & 217.65687658 &    59.723624 & M8 & 4   \\
Gl555      & 218.57030665 &   -12.519293 & M4 & 1   \\
2M1501+22  & 225.28405762 &    22.833836 & M9 & 4   \\
LHS3018    & 226.07657872 &    60.384562 & M1 & 1   \\
Gl581      & 229.86143531 &    -7.722272 & M3 & 1   \\
Gl585      & 230.96299622 &    17.465422 & M4 & 1   \\
LHS3080    & 232.97554057 &    28.852619 & M4 & 1   \\
Gl609      & 240.71170993 &    20.588769 & M4 & 1   \\
Gl625      & 246.35287094 &    54.304172 & M2 & 1   \\
Gl628      & 247.57513257 &   -12.662992 & M4 & 1   \\
LTT14949   & 250.20368457 &    36.316638 & M2 & 1   \\
GL643      & 253.85475375 &    -8.322944 & M4 & 1   \\
GJ1207     & 254.27401671 &    -4.349044 & M4 & 1   \\
LHS3262    & 255.84984701 &    51.406478 & M4 & 1   \\
GJ1209     & 256.09302479 &    16.931814 & M3 & 1   \\
Gl655      & 256.78099478 &    21.554112 & M3 & 1   \\
LTT15087   & 257.89507230 &    38.442849 & M4 & 1   \\
Gl678.1    & 262.59473243 &     5.548422 & M0 & 1   \\
Gl686      & 264.47272430 &    18.592155 & M1 & 1   \\
Gl694      & 265.98321814 &    43.378370 & M2 & 1   \\
GJ1223     & 270.69304642 &    37.517084 & M5 & 1   \\
Gl701      & 271.28155234 &    -3.031352 & M1 & 1   \\
Gl1227     & 275.61219859 &    62.049964 & M5 & 1   \\
Gl720B     & 278.86411519 &    45.761550 & M4 & 1   \\
Gl725B     & 280.69435149 &    59.627614 & M3 & 1   \\
Gl729      & 282.45607171 &   -23.836125 & M4 & 1   \\
Gl745A     & 286.77322036 &    20.887782 & M2 & 1   \\
Gl745B     & 286.80506716 &    20.876758 & M2 & 1   \\
G207-22    & 288.12267278 &    35.564413 & M2 & 1   \\
Gl752A     & 289.22996750 &     5.168470 & M3 & 1   \\
GJ1235     & 290.41104011 &    20.867176 & M4 & 1   \\
GJ1253     & 306.52229530 &    58.573236 & M5 & 1   \\
Gl809      & 313.33253189 &    62.153943 & M1 & 1   \\
LHS3713    & 327.06371612 &    27.927964 & M2 & 1   \\
Gl849      & 332.41848642 &    -4.640774 & M3 & 1   \\
Gl127-35   & 337.19158157 &    18.931698 & M0 & 1   \\
Gl867B     & 339.68865820 &   -20.614020 & M4 & 1   \\
Gl867AC    & 339.68997030 &   -20.620763 & M2 & 1   \\
Gl908      & 357.30266119 &     2.400989 & M1 & 1   \\
\enddata
\tablenotetext{a}{$\alpha$ and $\delta$ are reported in decimal degrees and J2000 coordinates.}
\tablenotetext{b}{1-PMSU \citep{1995AJ....110.1838R},
  2-DwarfArchives.org, 3-SDSS \citepalias{west10},
  4-\cite{2002AJ....123.2828C} }
\label{table:spec_types}
\end{deluxetable}
\end{center}

\begin{center}
\begin{deluxetable}{l|ll|ll}
\tablewidth{0pt}
\tabletypesize{\scriptsize}
\tablecaption{Absolute Magnitudes of SDSS M Dwarfs}
\tablehead{
\colhead{}&
\multicolumn{2}{c}{$\zeta = 1.1$} &
\multicolumn{2}{c}{$\zeta = 0.88$}
}

\startdata
r -z &   Active  &  Inactive & Active & Inactive \\     
\hline
1.05&  \nodata &   8.8       &    \nodata         &   \nodata          \\
1.35 &  \nodata &  9.1        &    \nodata         &    9.7         \\
1.65 & 8.4   &       9.8        &    9.3         &    10.4         \\
 1.95 &9.0    &  10.4    &      10.3       &     11.1        \\
 2.25 &10.3  &  11.3    &    11.1         &     12.0        \\
 2.55 &11.2  &  12.4    &    12.5         &    \nodata         \\
 2.85 &13.2  &   13.5    &    13.6         &   \nodata         \\
 3.15 &13.9  &   14.0    &     13.9        &    \nodata         \\
 3.45 &14.9  & \nodata    &   \nodata          &   \nodata          \\

\enddata
 \label{table:abs_mags}
\tablecomments{The typical uncertainty for the reported absolute
  magnitudes is $\sigma_{M_r} = 0.42$, which is directly computed from
  Equation \ref{eqn:sigk}.}
\end{deluxetable}
\end{center}

\subsection{Kinematics of M Dwarfs}\label{sec:kinematics}

In addition to an absolute magnitude estimate, the analysis returns
kinematic information for each subsample.  The interpretation of the
kinematic
results is complicated by the disjoint spatial volumes that 
the samples inhabit (see Table \ref{table:heights}).  We
discuss the mean motions of M dwarfs as traced by the solar reflex
motion, the velocity dispersions exhibited by active stars compared to 
their inactive counterparts and the
change in velocity dispersions seen in samples at different galactic
heights.

\subsubsection{Solar Peculiar Motion}
The statistical parallax method measures the 
reflex solar motion with respect to the mean velocity of the stellar subsample
being analyzed.  If subsamples possess
different mean velocities, this will be manifested as a change in the
solar reflex motion. 
For reference, we use a coordinate system with $U$ increasing towards the Galactic
center, $V$ increasing in the direction of solar motion, and $W$
increasing vertically upward (as in \citealp{Dehnen98}).  This system
is right--handed, with the angular momentum vector of the solar
orbital motion pointing towards the
South Galactic Pole.  

\begin{figure*}[htbp]
 \begin{center}$
\begin{array}{cc}  
  \includegraphics[scale=0.35]{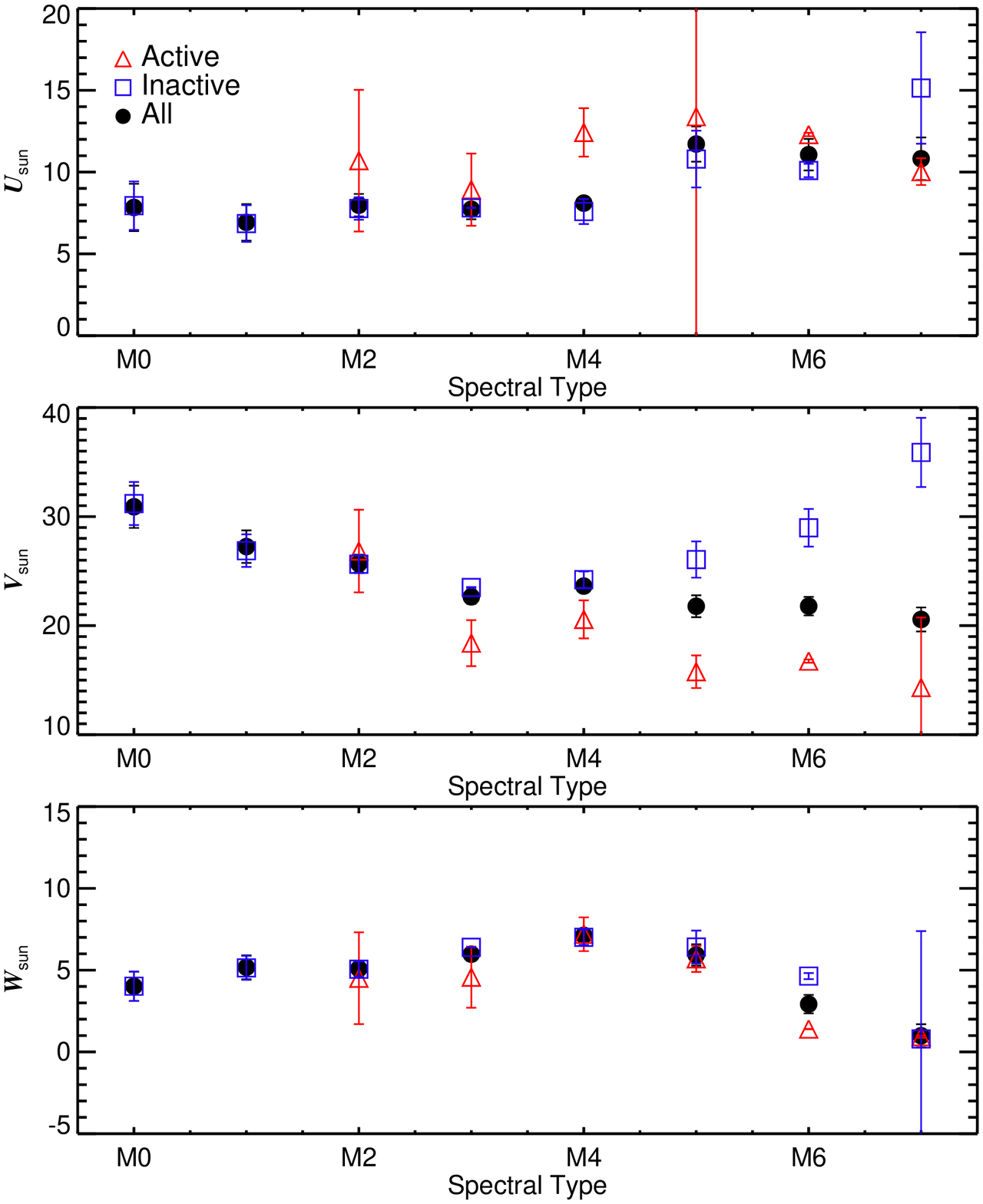}&
  \includegraphics[scale=0.35]{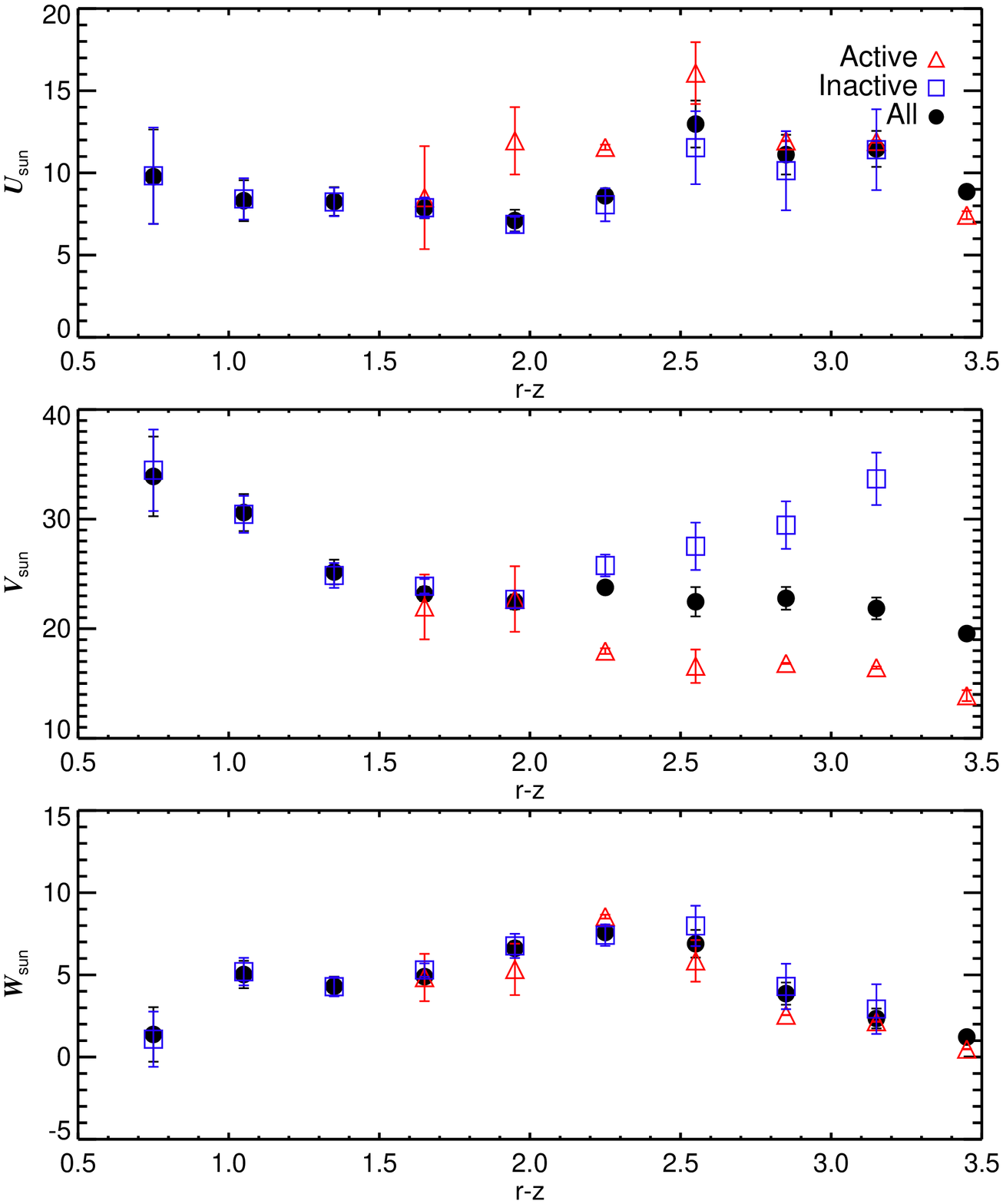} 
\end{array}$
\end{center}  
 \caption{Solar peculiar motion as a function of spectral type (left
   panel) and $r-z$ color (right panel).  
  The $V$ velocities exhibit a
 decline to later type (or redder) stars, likely due to an age
 effect, as shown by the separation between the active and inactive
 populations.
 The $W$ velocities exhibit a peaked behavior, while the $U$ velocities
 show a rise toward later types, which occurs at a bluer color in the
 active population.}
  \label{fig:solar_sp}
\end{figure*}

In the $U$ and $W$ directions, the solar reflex motion reflects the 
Sun's peculiar motion, as
these distributions should be centered on the local
standard of rest \citep[LSR;][]{2009AJ....137.4149F}.  The
solar reflex motion in the $V$ direction is a combination of the Sun's
peculiar motion and the asymmetric drift at the solar circle
\citep{1924ApJ....59..228S,1925ApJ....61..363S}.  This shifts the mean
$V$ velocity of a sample of typical disk-age M dwarfs, giving a $V$ reflex
motion larger than one measured from a population of young stars.  
Our results are shown in Figure
\ref{fig:solar_sp}, and we compare our peculiar solar velocities (i.e. the
negative of the reflex motion)
to previous work in Table \ref{table:old_uvws}.  Our values
were computed by taking the weighted mean across
all spectral types with the uncertainty given by the standard deviation.
While the $U$ and $W$ velocities remain relatively
constant with spectral type, the $V$ reflex motion is significantly larger at
early spectral types (M0-M2).  These stars are observed at larger distances
(Table \ref{table:heights}),
and also have larger velocity dispersions (as seen in Figure
\ref{fig:disps_sp_ipa}).    The same behavior is observed for the
bluest $r-z$ colors, which contain the same early--type M stars.  For
$r-z > 2$, the $V$ velocity of the Sun remains relatively
constant at 20 km s$^{-1}$.  This value compares favorably to the
$V$ velocities derived by \cite{1996AJ....112.2799H} and
\cite{2009AJ....137.4149F}, both measured with M dwarfs, but is 
larger than the V $\sim$ 5 km s$^{-1}$ value reported by
\cite{Dehnen98} from measurements of the Sun's motion through a very young
population, due to the age range of the stars that populate 
our subsamples.  At later types, where the separation in age is
most extreme between the active and inactive populations
\citep[see][and discussion in Sec 4.2.2 below]{2008AJ....135..785W},  the active stars in the 
sample show decreasing $V$ velocities, while the inactive stars
have increasing $V$ velocities, which supports the age hypothesis.

As mentioned above, the measured solar reflex velocity is the relative
motion between the Sun and the mean velocity of a subsample of M
dwarfs\footnote{We note that the solar motion has also been estimated
  by observing the direction and velocity of interstellar He entering
  the heliosphere \citep{2004A&A...426..835W}.}.  
Since the Sun's velocity is not changing, a measured change in the reflex
velocity
indicates a difference in the mean velocity of a particular subsample 
with respect
to the LSR.  While the $U$ and $W$ velocity dispersions are expected 
to be constant with
color (or spectral type), we find that there is significant structure in both
distributions.  The $W$ velocity distribution in Figure \ref{fig:solar_sp} 
displays a non-monotonic behavior, peaking near
$r-z \sim$ 2.3 (type M4).  This structure indicates the mean vertical
motion of the particular M dwarf subsample is varying, 
with both bluer and redder M dwarfs 
having a smaller mean velocity.
The $W$ velocities of active and inactive stars are not 
significantly different.  Meanwhile, the
$U$ velocity distribution shows a different behavior, remaining
relatively constant from M0-M5 
($r-z \sim 2.5$), 
and then exhibiting a rise toward
later types.  The active stars appear to show this rise in $U$ velocity
at a bluer color ($r-z \sim 2.0$), but then are joined by the inactive
stars at later types.
This may be due to the changes seen in the other ($V$, $W$) 
velocities and/or to the change in the absolute magnitude of the active
stars in bins where they are a significant fraction of the total
distribution.  Modeling of the populations to explore
changing the various parameters (activity fraction, absolute magnitude,
input velocities in each direction) is necessary in order to disentangle
these related effects.

\begin{center}
\begin{deluxetable*}{llll}
\tablewidth{0pt}
 \tablecaption{Solar Peculiar Motion}
 \tabletypesize{\small}
\tablehead{
 \colhead{Study} &
 \colhead{$U_{\odot}$\tablenotemark{a}} &
 \colhead{$V_{\odot}$\tablenotemark{a,b}} &
 \colhead{$W_{\odot}$\tablenotemark{a}}
}
 \startdata
\cite{1996AJ....112.2799H} & 9.1  $\pm$ 2 & 23.3 $\pm$ 2 & 7.6 $\pm$ 2\\
\cite{Dehnen98} & 10.00 $\pm$ 0.36  & 5.25 $\pm$ 0.62  & 7.17 $\pm$ 0.38   \\
\cite{Dehnen98}\tablenotemark{c} & 10 $\pm$ 1  & 22 $\pm$ 1  & 8 $\pm$ 2   \\
\cite{2009AJ....137.4149F} & 9 $\pm$ 1   &  20 $\pm$ 2 &  7 $\pm$ 1  \\
\cite{2009MNRAS.397.1286A} & 9.96 $\pm$ 0.33  & 5.25 $\pm$ 0.54   & 7.07 $\pm$ 0.34  \\
\cite{2010ApJ...716....1B}  &  10\tablenotemark{d}  &  20  & 6.5 $\pm$ 0.4\\
\cite{2010MNRAS.403.1829S} & 11.1 $\pm$ 1 & 12.24 $\pm$ 2 & 7.25 $\pm$0.5 \\
This Study\tablenotemark{e} & 8 $\pm$ 2   & 24 $\pm$ 3   &  7  $\pm$ 2\\
\enddata
\tablenotetext{a}{$UVW$ velocities are reported in km s$^{-1}$, with
  $U$ increasing radially inward, $V$ increasing in the direction of
  the Solar orbit, and $W$ increasing above the Galactic plane.}
\tablenotetext{b}{Measured relative to nearby M dwarfs, except for the
  \cite{Dehnen98} and \cite{2009MNRAS.397.1286A} studies.}
\tablenotetext{c}{Estimated from reddest bin in Figure 3 of \cite{Dehnen98}.}
\tablenotetext{d}{Assumed.}
\tablenotetext{e}{Error-weighted mean over spectral types M0-M7.}
 \label{table:old_uvws}
\end{deluxetable*}
\end{center}

\subsubsection{Velocity Dispersions}
\begin{figure*}[htbp]
%\centering
\begin{center}$
\begin{array}{cc} 
\includegraphics[scale=0.4]{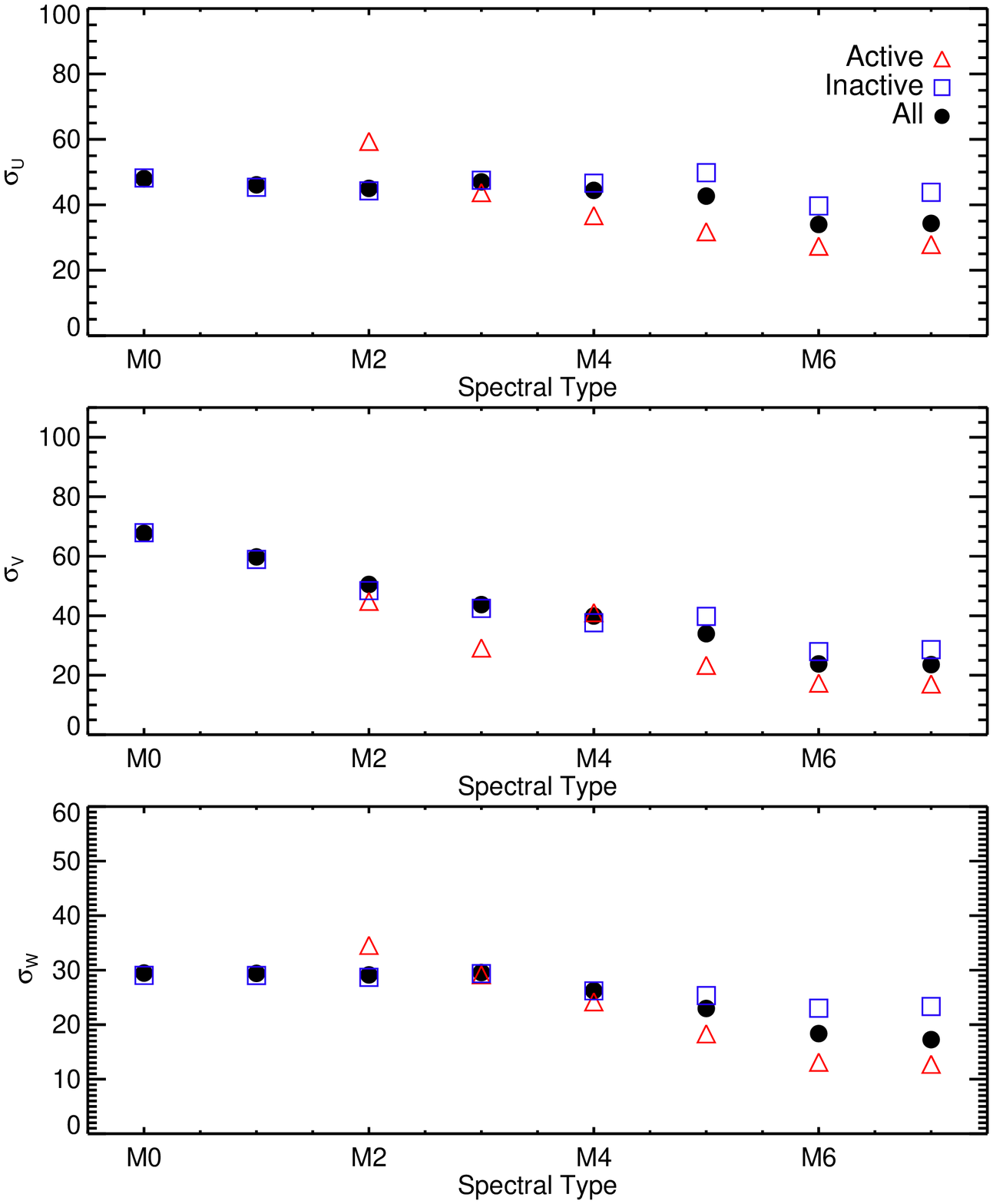} 
\includegraphics[scale=0.4]{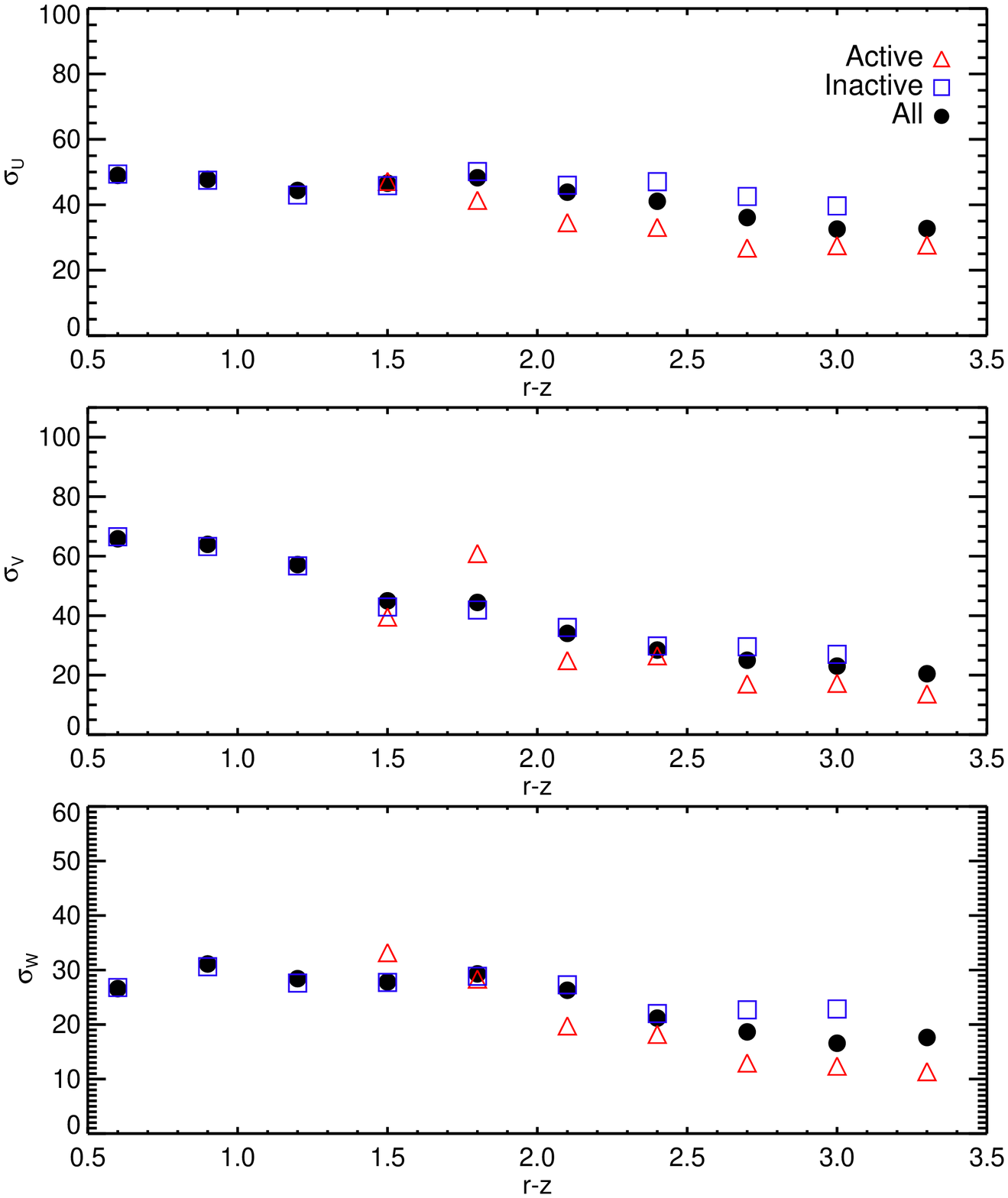}
\end{array}$
\end{center}
 \caption{$UVW$ velocity dispersions vs. spectral type (left panel)
   and color (right panel) for active
   (red triangles), inactive (blue squares) and all stars (filled
   black circles).  Active stars possess smaller
   dispersions at mid-late spectral types (red colors).  
   The M2 bin shows a strong vertex deviation so the velocity dispersions
   should not be interpreted as measurements along the $UVW$ axes.}
  \label{fig:disps_act}
\end{figure*}

The active M dwarfs are known to form a kinematically
colder population than their inactive counterparts \citep[e.g.,][]{1977A&A....60..263W,
  1996AJ....112.2799H}, as evidenced by smaller velocity dispersions.
This is usually interpreted
as an age effect, where younger M dwarfs have had less
time to interact with giant molecular clouds and spiral density
waves.  The velocity dispersion is predicted to vary as the square--root of 
age \citep[e.g.,][]{2001ASPC..228..235F,2002MNRAS.337..731H}.
Young stars also harbor strong magnetic fields that heat their chromospheres, 
allowing activity to be used as a proxy for age
\citep{1970MNRAS.148..463W}. Activity in low--mass stars has been shown to
depend on both spectral type and age; inactive, late--type M dwarfs are usually
much older than inactive early--type M dwarfs \citep{2008AJ....135..785W}.  
In  Table  \ref{table:disps_act},
we list the $UVW$ velocity dispersions for the active
and inactive subsamples by spectral type and color.  Figure \ref{fig:disps_act}
shows the results.  As expected, the dispersions of
active stars separate from their inactive counterparts at the same
spectral type (color) for mid-late M dwarfs.  In the first
bin where there are enough active stars to carry out the statistical
parallax analysis (at spectral type M2), 
the active stars have larger reported dispersions than the inactive ones.
The velocity ellipsoid in this case 
shows a strong vertex deviation, indicating 
a young population, which has not undergone significant dynamical heating
\citep{1998gaas.book.....B}, and the velocity dispersions should not be
interpreted as measurements along the $UVW$ axes.
The active M2 stars being members of a young population is 
consistent with age--activity relations that predict short activity
lifetimes for early--type M stars.
\citep{2008AJ....135..785W}.

\begin{center}
\begin{deluxetable*}{c|rrr|rrr|rrr}
\tablewidth{0pt}
 \tablecaption{Velocity Dispersions of M dwarfs}
 \tabletypesize{\scriptsize}

 \tablehead{
 \colhead{Spectral Type} &
 \multicolumn{3}{c}{$\sigma_U$ (km s$^{-1}$)}&
\multicolumn{3}{c}{$\sigma_V$ (km s$^{-1}$)}&
\multicolumn{3}{c}{$\sigma_W$ (km s$^{-1}$)}
}
 \startdata
	& 	Active	& 	Inactive 	& 	All &	Active	& 	Inactive 	& 	All &	Active	& 	Inactive 	& 	All \\
\hline
M0 &   \nodata  &     48  &     47  &     \nodata  &     67  &     67  &     \nodata  &     29  &     29 \\
M1 &  \nodata  &     45  &     46  &     \nodata  &     58  &     59  &     \nodata  &     29  &     29 \\
M2 &   59  &     44  &     45  &     44  &     66  &     48  &     34  &     28  &     29 \\
M3 &   43  &     47  &     47  &     29  &     51  &     42  &     29  &     29  &     29 \\
M4 &   36  &     46  &     44  &     41  &     41  &     37  &     24  &     26  &     26 \\
M5 &   31  &     49  &     42  &     23  &     39  &     39  &     18  &     25  &     22 \\
M6 &   27  &     39  &     33  &     17  &     28  &     27  &     13  &     23  &     18 \\
M7 &   27  &     43  &     34  &     17  &     29  &     28  &     12  &     23  &     17 \\

$r-z$&            & & &   & & &  && \\
\hline
0.75  &     \nodata  & 49       & 48  &   \nodata  &  66   &  65   &\nodata & 26  & 26\\
1.05  &     \nodata  &  47        & 47  &    \nodata & 63    &63   &   \nodata   &  30 & 31  \\
1.35  &     \nodata  &  42        & 44  &   \nodata  & 56    & 57  &      \nodata    & 27  & 28 \\
1.65  &     47            & 45        & 46  &     39         & 42   & 44  &      33   & 27  & 27  \\
1.95  &     41             & 50       & 48  &     60        &  41  & 44  &      28    & 28  & 29 \\
2.25  &     34             &  45        &  43 &     24        & 36  & 34  &      19   & 27  & 26  \\
2.55  &     33              &  47       & 41  &     26       & 29  & 28  &       18   & 22  & 21  \\
2.85  &     26               &  42         &  36 &     17     & 29   & 25  &      12   & 22  & 18 \\
3.15 &     27                &   39          & 32  &     17     & 26   &  23 &      12    & 22  & 16 \\
3.45 &    27                  & \nodata    & 32  &     13     &  \nodata  & 20  &      11    & \nodata  & 17  \\

\enddata
 \label{table:disps_act}
\tablecomments{The typical uncertainty for the dispersions is 5-10 km s$^{-1}$.
  Equation \ref{eqn:sigk}.}
\end{deluxetable*}
\end{center}

Another notable feature of Figure \ref{fig:disps_act} is the decline
in velocity dispersion at later types.  Since the type (color) bins
sample different spatial volumes (see Table \ref{table:heights}), this is likely
showing that stars that are located further above the plane of the Galaxy 
(in the early--type bins) are there because they are older and have 
undergone more dynamical interactions, giving correspondingly 
larger velocity dispersions.  To investigate this kinematic structure
in the Milky Way, we use the IPA parameter described in Equation \ref{eqn:ipa}
as a proxy for height.  

The mean height above the Plane, $Z$, can
be approximated as:
\begin{equation}
Z \approx d \sin b
\end{equation}
where $b$ is the mean Galactic latitude for the subsample ($\sim$
50$^{\circ}$ for SDSS as a whole).  This approximation neglects effects due to
the Sun's location above the plane \citep[15 pc; ][]{1995ApJ...444..874C}, but see
\citep{2010AJ....139.2679B} for a more precise expression.  Given the definition of IPA in
Equation \ref{eqn:ipa}, $Z$ and IPA are thus related by:
\begin{equation}
Z \approx {\rm IPA}~10^{(1-\frac{M_r}{5})}.  
\end{equation}
For reference, we tabulate various heights for a given IPA and $M_r$
in Table \ref{table:ipa_height}.  
To increase the number of stars in each IPA bin, we used stars at the
same absolute distance from
the Plane in both the northern and southern Galactic hemispheres, thus
implicitly assuming the Galaxy is symmetric about its mid-plane.

\begin{center}
\begin{deluxetable}{lllll}
\tablewidth{0pt}
 \tablecaption{Heights for various IPAs and Absolute Magnitudes\tablenotemark{a}\label{table:ipa_height}}
 \tabletypesize{\small}
 \tablehead{
 \colhead{IPA} &
 \multicolumn{3}{c}{$M_r$} 
}
 \startdata
    &   8  & 10 & 12 & 14\\
\hline
  100  &   25  &   10  &    3  &    1 \\
 1000  &  251  &  100  &   39  &   15 \\
 3000  &  753  &  300  &  119  &   47 \\
 5000  & 1255  &  500  &  199  &   79 \\
 6000  & 1507  &  600  &  238  &   95 \\
\enddata
\tablenotetext{a}{All heights reported in parsecs.}
\end{deluxetable}
\end{center}

\begin{figure*}[htbp]
\centering
\includegraphics[scale=0.3]{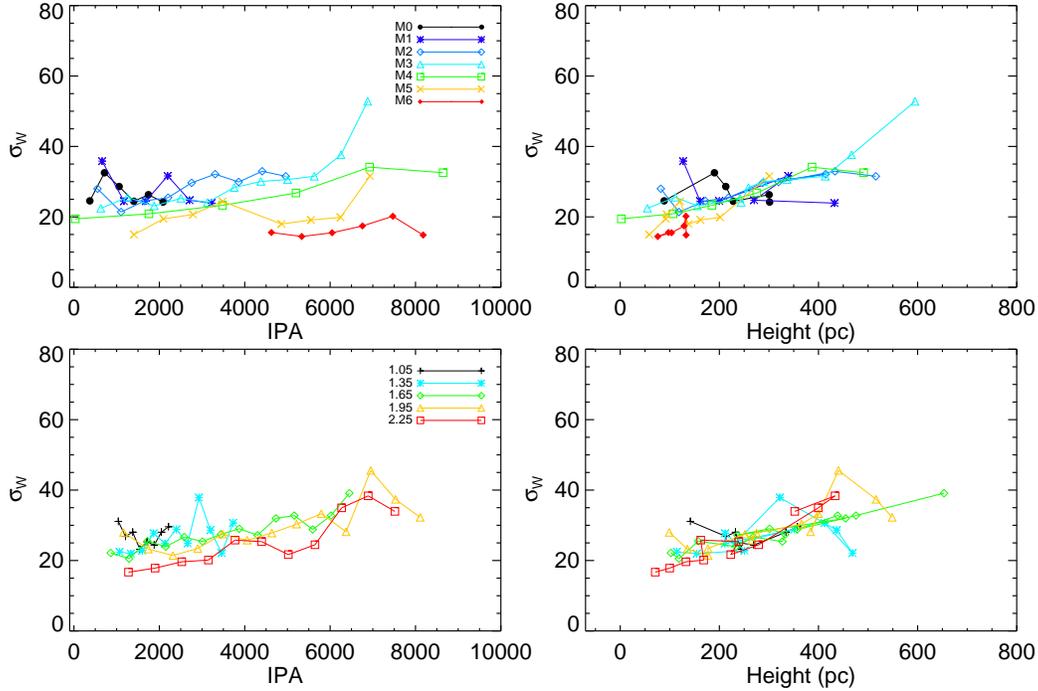} 
\caption{$W$ velocity dispersions vs. IPA (left panels) and height above the
  Plane (right panels).  The velocity dispersions increase with IPA/height
  but do not depend on color/spectral type.}
  \label{fig:disps_sp_ipa}
\end{figure*}

Previous kinematic
investigations \citep{2005AJ....130.1871B, 2007AJ....134.2418B,
  2008AJ....135..785W, 2009AJ....137.4149F,2010AJ....139.1808S} have
demonstrated an increase in velocity dispersion at larger Galactic
heights, rising from $(\sigma_U, \sigma_V, \sigma_W) \approx (30$ km
s$^{-1}$, $20$ km s$^{-1}$, $20$ km s$^{-1} )$ in the
local solar neighborhood, to  $(\sigma_U, \sigma_V, \sigma_W) \approx (50$ km
s$^{-1}$, $40$ km s$^{-1}$, $40$ km s$^{-1})$ at 1 kpc.
In Figure \ref{fig:disps_sp_ipa}, we plot the vertical $W$ velocity dispersion
as a function of IPA (left panels) and mean height above the Plane
(right panels), for spectral type (upper row) and color (lower row).
Recall that the advantage of the IPA is that no absolute magnitude is
assumed, while the heights are calculated using the absolute magnitude
from the statistical parallax solution for a given subsample.
The increase in velocity dispersion is seen in both height and IPA.
Furthermore, there is no strong dependence on spectral type or color,
as expected since the dynamical interactions should be controlled by 
the Galactic gravitational potential, with the stars (of any mass)
acting as collisionless test particles.  These results bridge the gap 
between the kinematic
investigations of \cite{2010ApJ...716....1B}, which examined stars
beyond our distances limits, and those examining nearby stars interior
to our study \citep{2004A&A...418..989N, 2002AJ....124.2721R}. 

The velocity dispersions for M dwarfs listed in Table \ref{table:disps_act} 
should be interpreted in light of these IPA/height results.
Earlier spectral types (i.e., M0 and M1) sample a larger range of distances 
and have a larger mean distance than later types 
(see Table \ref{table:heights}).  
Figure \ref{fig:disps_sp_ipa} shows that both early--type and late--type
M dwarfs have similar velocity dispersions when sampled at the
same Galactic height.

\section{Conclusions}\label{sec:conclusions}
%summarize results
We present a statistical parallax analysis of the most recent catalog of M
dwarfs identified with SDSS spectroscopy \citepalias{west10}.  Our sample
was subdivided on many criteria, to explore both the intrinsic changes
in low--mass dwarfs due to metallicity and magnetic activity, and to 
investigate their kinematics.  We have
demonstrated that $r-z$ color is a much better proxy for 
absolute magnitude than spectral type, and suggest that 
photometric parallaxes are the preferred method to determine
absolute magnitudes and 
distances for M dwarfs.

Some interesting trends with $M_r$ were revealed in our analysis.  
First, magnetically active
M dwarfs were shown to be intrinsically brighter at the same spectral
type or color than their inactive counterparts.  Eclipsing binary
studies have demonstrated that magnetic activity inflates a star's
radius \citep{2006Ap&SS.304...89R, 2007ApJ...660..732L, 2008A&A...478..507M,
2010ApJ...718..502M}, but the effect on effective
temperature and luminosity is less constrained.  Our results suggest
that activity may increase the radius {\it and} luminosity of
active low--mass stars.

Metallicity, which influences the luminosity of a star, was
also explored.  We divided the sample using the $\zeta$ parameter
\citep{2007ApJ...669.1235L} as a proxy for
metallicity.  Our results are similar to those
observed for higher mass stars:  low--metallicity M dwarfs are dimmer
at a given spectral type (or color) than high metallicity stars.  We
have quantified this effect for M dwarfs in the SDSS photometric
system.  

To isolate the effects of metallicity and activity, we separated
active and inactive stars for the same $\zeta$ as a function of color
and spectral type.  Activity still brightened stars at the same
$\zeta$, however the effect diminished at smaller $\zeta$ (lower
metallicity).

The statistical parallax analysis also allowed us to investigate 
the reflex solar motion and velocity dispersions for each subsample.  
The more distant, early--type stars, which are presumably older, 
have the largest reflex solar motion, particularly in the $V$ direction,
which we attribute to the increased asymmetric drift.  The active
and inactive stars exhibit expected behavior with the active star populations
having smaller mean motions relative to the Sun.
The inactive, late--type M dwarfs, which we identify as an older population
due to their lack of activity (ages $\gtrsim 4$ Gyrs,
\citealp{2008AJ....135..785W}) have velocity dispersions similar to
early--type M dwarfs, which we identify as old due to their greater 
vertical distance from the Galactic plane.   Thus, the activity and 
dynamical heating
age indicators give consistent results.  When the velocity dispersions
are analyzed as a function of vertical height or independent position 
altitude (IPA), all stars 
exhibit increasing dispersion at increasing height above the Plane.

As astronomy enters a new era of large photometric surveys, such as
PanSTARRS \citep{2002SPIE.4836..154K} and LSST \citep{2008arXiv0805.2366I}, it will be vital to develop
techniques for estimating the activity and metallicity of low--mass
stars from photometry alone.  There have already been efforts to
characterize metallicity using the color--color distributions of M
dwarfs in SDSS (\citealp{2009AIPC.1094..545L}; \citetalias{west10}), but those
data must be calibrated
with spectroscopic observations.  The work presented in this paper
highlights the need to determine these fundamental
parameters, since they affect the estimated distance to each star.
Finally, the importance of the dual
spectroscopic and photometric nature of SDSS cannot be overstated.
The large spectroscopic samples of M dwarfs it has acquired have
enabled many novel investigations, including this one.  Significant
spectroscopic followup of the next generation of surveys should
be a high priority.

We thank Adam Burgasser, Jacqueline Faherty, Rob Simcoe and Kevin
Covey for helpful discussions. We thank Neill Reid, Kelle Cruz and Richard Gray for making their
nearby stellar spectra available to us.  JJB personally acknowledges
Roy Halladay for inspiration and motivation throughout this work.  
JJB thanks the financial support of Adam Burgasser
and Kevin Luhman.  We also gratefully acknowledge the support of NSF grants AST 02-05875
and AST 06-07644 and NASA ADP grant NAG5-13111.   

Funding for the SDSS and SDSS-II has been provided by the Alfred P. Sloan Foundation, the Participating Institutions, the National Science Foundation, the U.S. Department of Energy, the National Aeronautics and Space Administration, the Japanese Monbukagakusho, the Max Planck Society, and the Higher Education Funding Council for England. The SDSS Web Site is http://www.sdss.org/.

The SDSS is managed by the Astrophysical Research Consortium for the Participating Institutions. The Participating Institutions are the American Museum of Natural History, Astrophysical Institute Potsdam, University of Basel, University of Cambridge, Case Western Reserve University, University of Chicago, Drexel University, Fermilab, the Institute for Advanced Study, the Japan Participation Group, Johns Hopkins University, the Joint Institute for Nuclear Astrophysics, the Kavli Institute for Particle Astrophysics and Cosmology, the Korean Scientist Group, the Chinese Academy of Sciences (LAMOST), Los Alamos National Laboratory, the Max-Planck-Institute for Astronomy (MPIA), the Max-Planck-Institute for Astrophysics (MPA), New Mexico State University, Ohio State University, University of Pittsburgh, University of Portsmouth, Princeton University, the United States Naval Observatory, and the University of Washington.


\begin{thebibliography}{dummy}

\bibitem[{{Abazajian} {et~al.}(2009)}]{2009ApJS..182..543A}
{Abazajian}, K.~N., {et~al.} 2009, \apjs, 182, 543

\bibitem[{{An} {et~al.}(2008)}]{2008ApJS..179..326A}
{An}, D., {et~al.} 2008, \apjs, 179, 326

\bibitem[{{Aumer} \& {Binney}(2009)}]{2009MNRAS.397.1286A}
{Aumer}, M., \& {Binney}, J.~J. 2009, \mnras, 397, 1286

\bibitem[{{Berger} {et~al.}(2006)}]{2006ApJ...644..475B}
{Berger}, D.~H., {et~al.} 2006, \apj, 644, 475

\bibitem[{{Bilir} {et~al.}(2009){Bilir}, {Karaali}, {Ak}, {Co{\c s}kuno{\u
  g}lu}, {Yaz}, \& {Cabrera-Lavers}}]{2009MNRAS.396.1589B}
{Bilir}, S., {Karaali}, S., {Ak}, S., {Co{\c s}kuno{\u g}lu}, K.~B., {Yaz}, E.,
  \& {Cabrera-Lavers}, A. 2009, \mnras, 396, 1589

\bibitem[{{Binney} \& {Merrifield}(1998)}]{1998gaas.book.....B}
{Binney}, J., \& {Merrifield}, M. 1998, {Galactic astronomy}, ed. {Binney,
  J.~\& Merrifield, M.}

\bibitem[{{Bochanski} {et~al.}(2010){Bochanski}, {Hawley}, {Covey}, {West},
  {Reid}, {Golimowski}, \& {Ivezi{\'c}}}]{2010AJ....139.2679B}
{Bochanski}, J.~J., {Hawley}, S.~L., {Covey}, K.~R., {West}, A.~A., {Reid},
  I.~N., {Golimowski}, D.~A., \& {Ivezi{\'c}}, {\v Z}. 2010, \aj, 139, 2679

\bibitem[{{Bochanski} {et~al.}(2005){Bochanski}, {Hawley}, {Reid}, {Covey},
  {West}, {Tinney}, \& {Gizis}}]{2005AJ....130.1871B}
{Bochanski}, J.~J., {Hawley}, S.~L., {Reid}, I.~N., {Covey}, K.~R., {West},
  A.~A., {Tinney}, C.~G., \& {Gizis}, J.~E. 2005, \aj, 130, 1871

\bibitem[{{Bochanski} {et~al.}(2007{\natexlab{a}}){Bochanski}, {Munn},
  {Hawley}, {West}, {Covey}, \& {Schneider}}]{2007AJ....134.2418B}
{Bochanski}, J.~J., {Munn}, J.~A., {Hawley}, S.~L., {West}, A.~A., {Covey},
  K.~R., \& {Schneider}, D.~P. 2007{\natexlab{a}}, \aj, 134, 2418

\bibitem[{{Bochanski} {et~al.}(2007{\natexlab{b}}){Bochanski}, {West},
  {Hawley}, \& {Covey}}]{2007AJ....133..531B}
{Bochanski}, J.~J., {West}, A.~A., {Hawley}, S.~L., \& {Covey}, K.~R.
  2007{\natexlab{b}}, \aj, 133, 531

\bibitem[{{Bochanski}(2008)}]{bochanskithesis}
{Bochanski}, Jr., J.~J. 2008, PhD thesis, University of Washington

\bibitem[{{Bond} {et~al.}(2010)}]{2010ApJ...716....1B}
{Bond}, N.~A., {et~al.} 2010, \apj, 716, 1

\bibitem[{{Chabrier} {et~al.}(2007){Chabrier}, {Gallardo}, \&
  {Baraffe}}]{2007A&A...472L..17C}
{Chabrier}, G., {Gallardo}, J., \& {Baraffe}, I. 2007, \aap, 472, L17

\bibitem[{{Clube} \& {Jones}(1971)}]{1971MNRAS.151..231C}
{Clube}, S.~V.~M., \& {Jones}, D.~H.~P. 1971, \mnras, 151, 231

\bibitem[{{Cohen}(1995)}]{1995ApJ...444..874C}
{Cohen}, M. 1995, \apj, 444, 874

\bibitem[{{Covey} {et~al.}(2008){Covey}, {Hawley}, {Bochanski}, {West}, {Reid},
  {Golimowski}, {Davenport}, {Henry}, {Uomoto}, \& {Holtzman}}]{covey08}
{Covey}, K.~R., {Hawley}, S.~L., {Bochanski}, J.~J., {West}, A.~A., {Reid},
  I.~N., {Golimowski}, D.~A., {Davenport}, J.~R.~A., {Henry}, T., {Uomoto}, A.,
  \& {Holtzman}, J.~A. 2008, \aj, 136, 1778

\bibitem[{{Covey} {et~al.}(2007)}]{2007AJ....134.2398C}
{Covey}, K.~R., {et~al.} 2007, \aj, 134, 2398

\bibitem[{{Cruz} \& {Reid}(2002)}]{2002AJ....123.2828C}
{Cruz}, K.~L., \& {Reid}, I.~N. 2002, \aj, 123, 2828

\bibitem[{{Dahn} {et~al.}(2002)}]{2002AJ....124.1170D}
{Dahn}, C.~C., {et~al.} 2002, \aj, 124, 1170

\bibitem[{{Daniels}(1978)}]{1978Daniels}
{Daniels}, R. 1978, {Introduction to Numerical Methods and Optimization
  Techniques} (New York: North-Holland)

\bibitem[{{Davenport} {et~al.}(2007){Davenport}, {Bochanski}, {Covey},
  {Hawley}, {West}, \& {Schneider}}]{2007AJ....134.2430D}
{Davenport}, J.~R.~A., {Bochanski}, J.~J., {Covey}, K.~R., {Hawley}, S.~L.,
  {West}, A.~A., \& {Schneider}, D.~P. 2007, \aj, 134, 2430

\bibitem[{{Dehnen} \& {Binney}(1998)}]{Dehnen98}
{Dehnen}, W., \& {Binney}, J.~J. 1998, \mnras, 298, 387

\bibitem[{{Dhital} {et~al.}(2010){Dhital}, {West}, {Stassun}, \&
  {Bochanski}}]{2010AJ....139.2566D}
{Dhital}, S., {West}, A.~A., {Stassun}, K.~G., \& {Bochanski}, J.~J. 2010, \aj,
  139, 2566

\bibitem[{{ESA}(1997)}]{1997yCat.1239....0E}
{ESA}. 1997, VizieR Online Data Catalog, 1239, 0

\bibitem[{{Fernley} {et~al.}(1998){Fernley}, {Barnes}, {Skillen}, {Hawley},
  {Hanley}, {Evans}, {Solano}, \& {Garrido}}]{1998A&A...330..515F}
{Fernley}, J., {Barnes}, T.~G., {Skillen}, I., {Hawley}, S.~L., {Hanley},
  C.~J., {Evans}, D.~W., {Solano}, E., \& {Garrido}, R. 1998, \aap, 330, 515

\bibitem[{{Fuchs} {et~al.}(2001){Fuchs}, {Dettbarn}, {Jahrei{\ss}}, \&
  {Wielen}}]{2001ASPC..228..235F}
{Fuchs}, B., {Dettbarn}, C., {Jahrei{\ss}}, H., \& {Wielen}, R. 2001, in
  Astronomical Society of the Pacific Conference Series, Vol. 228, Dynamics of
  Star Clusters and the Milky Way, ed. {S.~Deiters, B.~Fuchs, A.~Just,
  R.~Spurzem, \& R.~Wielen}, 235--+

\bibitem[{{Fuchs} {et~al.}(2009)}]{2009AJ....137.4149F}
{Fuchs}, B., {et~al.} 2009, \aj, 137, 4149

\bibitem[{{Fukugita} {et~al.}(1996){Fukugita}, {Ichikawa}, {Gunn}, {Doi},
  {Shimasaku}, \& {Schneider}}]{1996AJ....111.1748F}
{Fukugita}, M., {Ichikawa}, T., {Gunn}, J.~E., {Doi}, M., {Shimasaku}, K., \&
  {Schneider}, D.~P. 1996, \aj, 111, 1748

\bibitem[{{H{\"a}nninen} \& {Flynn}(2002)}]{2002MNRAS.337..731H}
{H{\"a}nninen}, J., \& {Flynn}, C. 2002, \mnras, 337, 731

\bibitem[{{Hawley} {et~al.}(1996){Hawley}, {Gizis}, \&
  {Reid}}]{1996AJ....112.2799H}
{Hawley}, S.~L., {Gizis}, J.~E., \& {Reid}, I.~N. 1996, \aj, 112, 2799

\bibitem[{{Hawley} {et~al.}(1986){Hawley}, {Jefferys}, {Barnes}, \&
  {Lai}}]{1986ApJ...302..626H}
{Hawley}, S.~L., {Jefferys}, W.~H., {Barnes}, III, T.~G., \& {Lai}, W. 1986,
  \apj, 302, 626

\bibitem[{{Hawley} {et~al.}(2002)}]{2002AJ....123.3409H}
{Hawley}, S.~L., {et~al.} 2002, \aj, 123, 3409

\bibitem[{{Ivezic} {et~al.}(2008)}]{2008arXiv0805.2366I}
{Ivezic}, Z., {et~al.} 2008, ArXiv e-prints

\bibitem[{{Ivezi{\'c}} {et~al.}(2008)}]{Ivezic08}
{Ivezi{\'c}}, Z., {et~al.} 2008, \apj, 684, 287

\bibitem[{{Johnson} \& {Apps}(2009)}]{2009ApJ...699..933J}
{Johnson}, J.~A., \& {Apps}, K. 2009, \apj, 699, 933

\bibitem[{{Juri{\'c}} {et~al.}(2008)}]{2008ApJ...673..864J}
{Juri{\'c}}, M., {et~al.} 2008, \apj, 673, 864

\bibitem[{{Kaiser} {et~al.}(2002)}]{2002SPIE.4836..154K}
{Kaiser}, N., {et~al.} 2002, in Proceedings of the SPIE, Volume 4836, pp.
  154-164 (2002)., ed. J.~A. {Tyson} \& S.~{Wolff}, Vol. 4836, 154--164

\bibitem[{{Kirkpatrick} {et~al.}(1991){Kirkpatrick}, {Henry}, \&
  {McCarthy}}]{1991ApJS...77..417K}
{Kirkpatrick}, J.~D., {Henry}, T.~J., \& {McCarthy}, D.~W. 1991, \apjs, 77, 417

\bibitem[{{Kraus} \& {Hillenbrand}(2007)}]{2007ApJ...662..413K}
{Kraus}, A.~L., \& {Hillenbrand}, L.~A. 2007, \apj, 662, 413

\bibitem[{{Laughlin} {et~al.}(1997){Laughlin}, {Bodenheimer}, \&
  {Adams}}]{1997ApJ...482..420L}
{Laughlin}, G., {Bodenheimer}, P., \& {Adams}, F.~C. 1997, \apj, 482, 420

\bibitem[{{Layden} {et~al.}(1996){Layden}, {Hanson}, {Hawley}, {Klemola}, \&
  {Hanley}}]{1996AJ....112.2110L}
{Layden}, A.~C., {Hanson}, R.~B., {Hawley}, S.~L., {Klemola}, A.~R., \&
  {Hanley}, C.~J. 1996, \aj, 112, 2110

\bibitem[{{L{\'e}pine}(2009)}]{2009AIPC.1094..545L}
{L{\'e}pine}, S. 2009, in American Institute of Physics Conference Series, Vol.
  1094, American Institute of Physics Conference Series, ed. {E.~Stempels},
  545--548

\bibitem[{{L{\'e}pine} {et~al.}(2007){L{\'e}pine}, {Rich}, \&
  {Shara}}]{2007ApJ...669.1235L}
{L{\'e}pine}, S., {Rich}, R.~M., \& {Shara}, M.~M. 2007, \apj, 669, 1235

\bibitem[{{L{\'o}pez-Morales}(2007)}]{2007ApJ...660..732L}
{L{\'o}pez-Morales}, M. 2007, \apj, 660, 732

\bibitem[{{Luyten}(1925)}]{1925ApJ....62....8L}
{Luyten}, W.~J. 1925, \apj, 62, 8

\bibitem[{{Malmquist}(1936)}]{m36}
{Malmquist}, K.~G. 1936, Stockholm Obs. Medd., 26

\bibitem[{{Mohanty} {et~al.}(2009){Mohanty}, {Stassun}, \&
  {Mathieu}}]{2009ApJ...697..713M}
{Mohanty}, S., {Stassun}, K.~G., \& {Mathieu}, R.~D. 2009, \apj, 697, 713

\bibitem[{{Morales} {et~al.}(2010){Morales}, {Gallardo}, {Ribas}, {Jordi},
  {Baraffe}, \& {Chabrier}}]{2010ApJ...718..502M}
{Morales}, J.~C., {Gallardo}, J., {Ribas}, I., {Jordi}, C., {Baraffe}, I., \&
  {Chabrier}, G. 2010, \apj, 718, 502

\bibitem[{{Morales} {et~al.}(2008){Morales}, {Ribas}, \&
  {Jordi}}]{2008A&A...478..507M}
{Morales}, J.~C., {Ribas}, I., \& {Jordi}, C. 2008, \aap, 478, 507

\bibitem[{{Mullan} \& {MacDonald}(2001)}]{2001ApJ...559..353M}
{Mullan}, D.~J., \& {MacDonald}, J. 2001, \apj, 559, 353

\bibitem[{{Munn} {et~al.}(2004)}]{2004AJ....127.3034M}
{Munn}, J.~A., {et~al.} 2004, \aj, 127, 3034

\bibitem[{{Munn} {et~al.}(2008)}]{2008AJ....136..895M}
---. 2008, \aj, 136, 895

\bibitem[{{Murray}(1983)}]{1983veas.book.....M}
{Murray}, C.~A. 1983, {Vectorial astrometry}, ed. {Murray, C.~A.}

\bibitem[{Nelder \& Mead(1965)}]{optimizationsimplex_nelder_1965}
Nelder, J.~A., \& Mead, R. 1965, The Computer Journal, 7, 308

\bibitem[{{Nordstr{\"o}m} {et~al.}(2004){Nordstr{\"o}m}, {Mayor}, {Andersen},
  {Holmberg}, {Pont}, {J{\o}rgensen}, {Olsen}, {Udry}, \&
  {Mowlavi}}]{2004A&A...418..989N}
{Nordstr{\"o}m}, B., {Mayor}, M., {Andersen}, J., {Holmberg}, J., {Pont}, F.,
  {J{\o}rgensen}, B.~R., {Olsen}, E.~H., {Udry}, S., \& {Mowlavi}, N. 2004,
  \aap, 418, 989

\bibitem[{{Perryman} {et~al.}(2001){Perryman}, {de Boer}, {Gilmore}, {H{\o}g},
  {Lattanzi}, {Lindegren}, {Luri}, {Mignard}, {Pace}, \& {de
  Zeeuw}}]{2001A&A...369..339P}
{Perryman}, M.~A.~C., {de Boer}, K.~S., {Gilmore}, G., {H{\o}g}, E.,
  {Lattanzi}, M.~G., {Lindegren}, L., {Luri}, X., {Mignard}, F., {Pace}, O., \&
  {de Zeeuw}, P.~T. 2001, \aap, 369, 339

\bibitem[{{Pier} {et~al.}(2003){Pier}, {Munn}, {Hindsley}, {Hennessy}, {Kent},
  {Lupton}, \& {Ivezi{\'c}}}]{2003AJ....125.1559P}
{Pier}, J.~R., {Munn}, J.~A., {Hindsley}, R.~B., {Hennessy}, G.~S., {Kent},
  S.~M., {Lupton}, R.~H., \& {Ivezi{\'c}}, {\v Z}. 2003, \aj, 125, 1559

\bibitem[{{Popowski} \& {Gould}(1998)}]{1998ApJ...506..259P}
{Popowski}, P., \& {Gould}, A. 1998, \apj, 506, 259

\bibitem[{{Reid}(1997)}]{1997AJ....114..161R}
{Reid}, I.~N. 1997, \aj, 114, 161

\bibitem[{{Reid} {et~al.}(1997){Reid}, {Gizis}, {Cohen}, {Pahre}, {Hogg},
  {Cowie}, {Hu}, \& {Songaila}}]{1997PASP..109..559R}
{Reid}, I.~N., {Gizis}, J.~E., {Cohen}, J.~G., {Pahre}, M.~A., {Hogg}, D.~W.,
  {Cowie}, L., {Hu}, E., \& {Songaila}, A. 1997, \pasp, 109, 559

\bibitem[{{Reid} {et~al.}(2002){Reid}, {Gizis}, \&
  {Hawley}}]{2002AJ....124.2721R}
{Reid}, I.~N., {Gizis}, J.~E., \& {Hawley}, S.~L. 2002, \aj, 124, 2721

\bibitem[{{Reid} \& {Hawley}(2005)}]{2005nlds.book.....R}
{Reid}, I.~N., \& {Hawley}, S.~L. 2005, {New light on dark stars : red dwarfs,
  low-mass stars, brown dwarfs} (New Light on Dark Stars Red Dwarfs, Low-Mass
  Stars, Brown Stars, by I.N.~Reid and S.L.~Hawley.~ Springer-Praxis books in
  astrophysics and astronomy.~Praxis Publishing Ltd, 2005.~ ISBN 3-540-25124-3)

\bibitem[{{Reid} {et~al.}(1995){Reid}, {Hawley}, \&
  {Gizis}}]{1995AJ....110.1838R}
{Reid}, I.~N., {Hawley}, S.~L., \& {Gizis}, J.~E. 1995, \aj, 110, 1838

\bibitem[{{Reid} \& {Majewski}(1993)}]{1993ApJ...409..635R}
{Reid}, N., \& {Majewski}, S.~R. 1993, \apj, 409, 635

\bibitem[{{Ribas}(2006)}]{2006Ap&SS.304...89R}
{Ribas}, I. 2006, \apss, 304, 89

\bibitem[{{Richards} {et~al.}(2002)}]{2002AJ....123.2945R}
{Richards}, G.~T., {et~al.} 2002, \aj, 123, 2945

\bibitem[{{Riedel} {et~al.}(2010)}]{2010arXiv1008.0648R}
{Riedel}, A.~R., {et~al.} 2010, ArXiv e-prints

\bibitem[{{Rojas-Ayala} {et~al.}(2010){Rojas-Ayala}, {Covey}, {Muirhead}, \&
  {Lloyd}}]{2010ApJ...720L.113R}
{Rojas-Ayala}, B., {Covey}, K.~R., {Muirhead}, P.~S., \& {Lloyd}, J.~P. 2010,
  \apjl, 720, L113

\bibitem[{{Sandage} \& {Eggen}(1959)}]{1959MNRAS.119..278S}
{Sandage}, A.~R., \& {Eggen}, O.~J. 1959, \mnras, 119, 278

\bibitem[{{Schlegel} {et~al.}(1998){Schlegel}, {Finkbeiner}, \&
  {Davis}}]{1998ApJ...500..525S}
{Schlegel}, D.~J., {Finkbeiner}, D.~P., \& {Davis}, M. 1998, \apj, 500, 525

\bibitem[{{Schmidt} {et~al.}(2010){Schmidt}, {West}, {Hawley}, \&
  {Pineda}}]{2010AJ....139.1808S}
{Schmidt}, S.~J., {West}, A.~A., {Hawley}, S.~L., \& {Pineda}, J.~S. 2010, \aj,
  139, 1808

\bibitem[{{Sch{\"o}nrich} {et~al.}(2010){Sch{\"o}nrich}, {Binney}, \&
  {Dehnen}}]{2010MNRAS.403.1829S}
{Sch{\"o}nrich}, R., {Binney}, J., \& {Dehnen}, W. 2010, \mnras, 403, 1829

\bibitem[{{Sesar} {et~al.}(2008){Sesar}, {Ivezi{\'c}}, \&
  {Juri{\'c}}}]{Sesar08}
{Sesar}, B., {Ivezi{\'c}}, {\v Z}., \& {Juri{\'c}}, M. 2008, \apj, 689, 1244

\bibitem[{{Skrutskie} {et~al.}(2006)}]{2006AJ....131.1163S}
{Skrutskie}, M.~F., {et~al.} 2006, \aj, 131, 1163

\bibitem[{{Stelzer} {et~al.}(2010){Stelzer}, {Scholz}, {Argiroffi}, \&
  {Micela}}]{2010MNRAS.tmp.1180S}
{Stelzer}, B., {Scholz}, A., {Argiroffi}, C., \& {Micela}, G. 2010, \mnras,
  1180

\bibitem[{{Strauss} {et~al.}(2002)}]{2002AJ....124.1810S}
{Strauss}, M.~A., {et~al.} 2002, \aj, 124, 1810

\bibitem[{{Str{\"o}mberg}(1924)}]{1924ApJ....59..228S}
{Str{\"o}mberg}, G. 1924, \apj, 59, 228

\bibitem[{{Str{\"o}mberg}(1925)}]{1925ApJ....61..363S}
---. 1925, \apj, 61, 363

\bibitem[{{Strugnell} {et~al.}(1986){Strugnell}, {Reid}, \&
  {Murray}}]{1986MNRAS.220..413S}
{Strugnell}, P., {Reid}, N., \& {Murray}, C.~A. 1986, \mnras, 220, 413

\bibitem[{{van Herk}(1965)}]{1965BAN....18...71V}
{van Herk}, G. 1965, \bain, 18, 71

\bibitem[{{van Leeuwen}(2007)}]{2007A&A...474..653V}
{van Leeuwen}, F. 2007, \aap, 474, 653

\bibitem[{{Vrba} {et~al.}(2004)}]{2004AJ....127.2948V}
{Vrba}, F.~J., {et~al.} 2004, \aj, 127, 2948

\bibitem[{{West} {et~al.}(2008){West}, {Hawley}, {Bochanski}, {Covey}, {Reid},
  {Dhital}, {Hilton}, \& {Masuda}}]{2008AJ....135..785W}
{West}, A.~A., {Hawley}, S.~L., {Bochanski}, J.~J., {Covey}, K.~R., {Reid},
  I.~N., {Dhital}, S., {Hilton}, E.~J., \& {Masuda}, M. 2008, \aj, 135, 785

\bibitem[{{West} {et~al.}(2005){West}, {Walkowicz}, \&
  {Hawley}}]{2005PASP..117..706W}
{West}, A.~A., {Walkowicz}, L.~M., \& {Hawley}, S.~L. 2005, \pasp, 117, 706

\bibitem[{{West} {et~al.}(2011)}]{west10}
{West}, A.~A., {et~al.} 2011, {AJ, submitted}

\bibitem[{{Wielen}(1977)}]{1977A&A....60..263W}
{Wielen}, R. 1977, \aap, 60, 263

\bibitem[{{Wilson} \& {Woolley}(1970)}]{1970MNRAS.148..463W}
{Wilson}, O., \& {Woolley}, R. 1970, \mnras, 148, 463

\bibitem[{{Wilson} {et~al.}(1991){Wilson}, {Barnes}, {Hawley}, \&
  {Jefferys}}]{1991ApJ...378..708W}
{Wilson}, T.~D., {Barnes}, III, T.~G., {Hawley}, S.~L., \& {Jefferys}, W.~H.
  1991, \apj, 378, 708

\bibitem[{{Witte}(2004)}]{2004A&A...426..835W}
{Witte}, M. 2004, \aap, 426, 835

\bibitem[{{Woolf} {et~al.}(2009){Woolf}, {L{\'e}pine}, \&
  {Wallerstein}}]{2009PASP..121..117W}
{Woolf}, V.~M., {L{\'e}pine}, S., \& {Wallerstein}, G. 2009, \pasp, 121, 117

\bibitem[{{York} {et~al.}(2000)}]{2000AJ....120.1579Y}
{York}, D.~G., {et~al.} 2000, \aj, 120, 1579

\end{thebibliography}
\end{document}